%% file: Cooperative_Spectrum_Sensing---main.tex
\documentclass[journal,twoside]{IEEEtran}

\ifCLASSOPTIONcompsoc
\else
\fi

\ifCLASSINFOpdf
\else
\fi

\usepackage{bm}
\usepackage{amsmath}
\usepackage{amsfonts}
\usepackage{amssymb}
\usepackage{graphicx}
\usepackage[tight,footnotesize]{subfigure}
\usepackage{url}
\usepackage{cite}

\usepackage[dvipsnames,usenames]{color}

\def\CN{{\cal N}_{\mathbb C}}
\def\Ex{{\mathbb E}}
\def\a{{\bm a}}

\def\h{{\bm h}}
\def\i{{\bm i}}

\def\v{{\bm v}}
\def\w{{\bm w}}
\def\x{{\bm x}}
\def\y{{\bm y}}

\def\A{{\bm A}}
\def\B{{\bm B}}
\def\C{{\bm C}}
\def\D{{\bm D}}
\def\E{{\bm E}}
\def\F{{\bm F}}
\def\G{{\bm G}}
\def\H{{\bm H}}
\def\I{{\bm I}}

\def\M{{\bm M}}
\def\O{{\bm O}}
\def\P{{\bm P}}
\def\Q{{\bm Q}}
\def\R{{\bm R}}
\def\S{{\bm S}}
\def\U{{\bm U}}
\def\V{{\bm V}}
\def\W{{\bm W}}

\def\Z{{\bm Z}}
\def\0{{\bm 0}}
\def\1{{\bm 1}}

\def\mub{{\bm \mu}}

\def\Sigmab{{\bm \Sigma}}

\begin{document}

\title{Orthogonality and Cooperation in Collaborative Spectrum Sensing through MIMO Decision Fusion}

\author{Pierluigi~Salvo~Rossi,~\IEEEmembership{Senior~Member,~IEEE}, Domenico~Ciuonzo,~\IEEEmembership{Student~Member,~IEEE}, \\ and Gianmarco~Romano,~\IEEEmembership{Member,~IEEE}
\thanks{Manuscript received February 12, 2013; revised June 14, 2013 and July 30, 2013; accepted August 12, 2013. 
The associate editor coordinating the review process of this paper and approving it for publication was Dr. G. Abreu.}
\thanks{This work was supported in part by the project ``Embedded Systems'' within the research framework Networks of Excellence, POR Campania FSE 2007/2013, Italy.}
\thanks{The authors are with the Department of Industrial and Information Engineering, Second University of Naples, Aversa (CE), Italy. Email: \{salvorossi,~domenico.ciuonzo,~gianmarco.romano\}@ieee.org}
\thanks{Digital Object Identifier}
}

\markboth{IEEE Transactions on Wireless Communications,~Vol.~*, No.~*, Month~yyyy}%
{Salvo Rossi \MakeLowercase{\textit{et al.}}: Orthogonality and Cooperation in Collaborative Spectrum Sensing through MIMO Decision Fusion}

\maketitle

\begin{abstract}
This paper deals with spectrum sensing for cognitive radio scenarios where the decision fusion center (DFC) exploits array processing.
More specifically, we explore the impact of user cooperation and orthogonal transmissions among secondary users (SUs) on the reporting channel.
To this aim four protocols are considered: (i) non-orthogonal and non-cooperative; (ii) orthogonal and non-cooperative; (iii) non-orthogonal and cooperative; (iv) orthogonal and cooperative.
The DFC employs maximum ratio combining (MRC) rule and performance are evaluated in terms of complementary receiver operating characteristic (CROC).
Analytical results, coupled with Monte Carlo simulations, are presented.
\end{abstract}


\input{Cooperative_Spectrum_Sensing---sec1.tex}

\input{Cooperative_Spectrum_Sensing---sec2.tex}

\input{Cooperative_Spectrum_Sensing---sec3.tex}
\input{Cooperative_Spectrum_Sensing---sec4.tex}

\section{Conclusion}
\label{sec:conclusion}
Focusing on a cognitive radio framework, in this paper we have considered spectrum sensing through MIMO decision fusion.
The aim was to assess the impact of user cooperation and orthogonal transmissions within this context. 
For that reason four protocols have been considered and compared for transmitting local decisions from the secondary users to the decision fusion center, namely MAC, PAC, CMAC and CPAC.
The impact of user cooperation is apparent when comparing the CROC (obtained through both Monte Carlo simulations and theoretical results) of the protocols: CMAC (resp. CPAC) improves MAC (resp. PAC) performance significantly. 
Also, the impact of orthogonality is significant when user cooperation is not considered (PAC improves MAC) while negligible when user cooperation is adopted (CMAC performs similar to CPAC).
Therefore user cooperation must be considered as an appealing technique to be employed on the reporting channel in spectrum sensing contexts.

\section*{Acknowledgment}
The authors would like to thank the Associate Editor and the anonymous reviewers for dedicating their time to the final version of this manuscript.

\input{Cooperative_Spectrum_Sensing---secAPP.tex}



\begin{IEEEbiography}[{\includegraphics[width=3in,height=1.3in,clip,keepaspectratio]{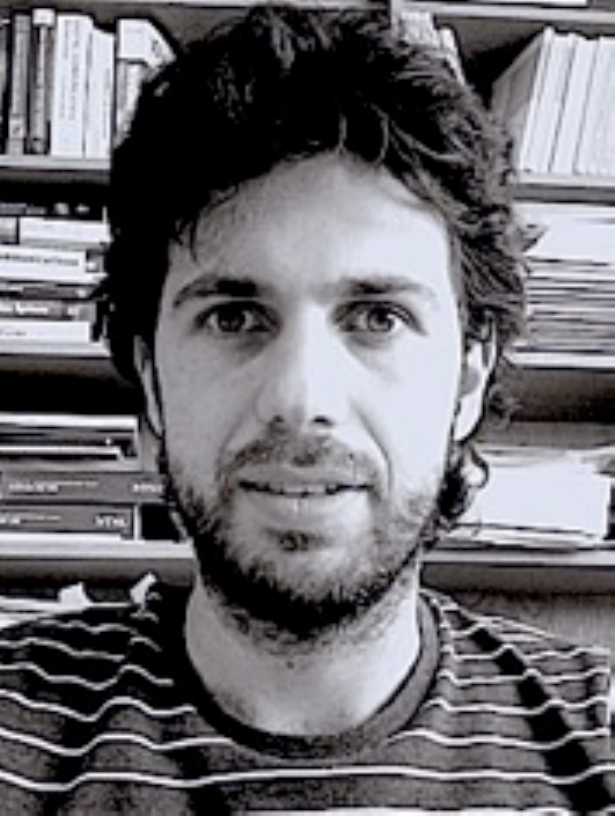}}]{Pierluigi~Salvo~Rossi (SM'11)}
was born in Naples, Italy, on April 26, 1977. 
He received the ``Laurea'' degree in telecommunications engineering (\emph{summa cum laude}) and the Ph.D. degree in computer engineering in 2002 and 2005, respectively, both from the University of Naples ``Federico~II'', Naples, Italy. 
During his Ph.D. studies, he has been a researcher at the Inter-departmental Research Center for Signals Analysis and Synthesis (CIRASS), University of Naples ``Federico~II'', Naples, Italy; 
at the Department of Information Engineering, Second University of Naples, Aversa (CE), Italy; 
at the the Communications and Signal Processing Laboratory (CSPL), Department of Electrical and Computer Engineering, Drexel University, Philadelphia, PA, US.; 
and an adjunct professor at the Faculty of Engineering, Second University of Naples, Aversa (CE), Italy. 
From 2005 to 2008 he worked as a postdoc at the Department of Computer Science and Systems, University of Naples ``Federico~II'', Naples, Italy; 
at the Department of Information Engineering, Second University of Naples, Aversa (CE), Italy; 
at the Department of Electronics and Telecommunications, Norwegian University of Science and Technology, Trondheim, Norway; 
and visited the Department of Electrical and Information Technology, Lund University, Lund, Sweden. 
Since 2008, he is an assistant professor (tenured in 2011) in telecommunications at the Department of Industrial and Information Engineering, Second University of Naples, Aversa (CE), Italy; and has been a guest researcher at the Department of Electronics and Telecommunications, Norwegian University of Science and Technology, Trondheim, Norway. 
He is an Editor for the {\sc IEEE Communications Letters}.
His research interests fall within the areas of communications and signal processing.
\end{IEEEbiography}

\vfill
\newpage

\begin{IEEEbiography}[{\includegraphics[width=3in,height=1.3in,clip,keepaspectratio]{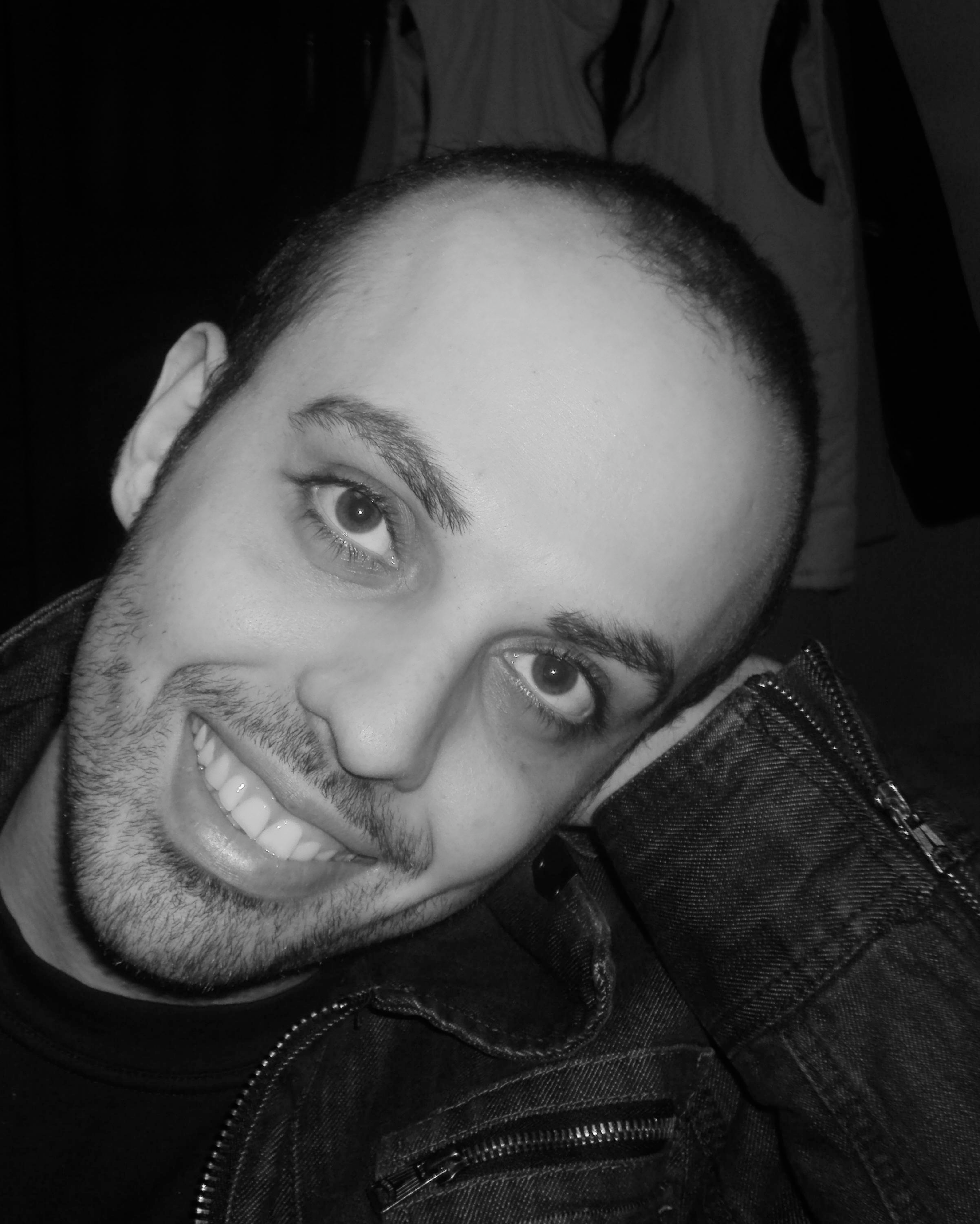}}]{Domenico~Ciuonzo (S'11)}
was born in Aversa (CE), Italy, on June 29th, 1985. 
He received the B.Sc. (\emph{summa cum laude}), the M.Sc. (\emph{summa cum laude}) degrees in computer engineering and the Ph.D. in electronic engineering, respectively in 2007, 2009 and 2013, from the Second University of Naples, Aversa (CE), Italy.
In 2011 he was involved in the Visiting Researcher Programme of the former NATO Underwater Research Center (now Centre for Maritime Research and Experimentation), La Spezia, Italy; he worked in the "Maritime Situation Awareness" project. 
In 2012 he was a visiting scholar at the Electrical and Computer Engineering Department, University of Connecticut, Storrs, US.  
His research interests are mainly in the areas of data and decision fusion, statistical signal processing, target tracking and probabilistic graphical models.  
Dr.~Ciuonzo is a reviewer for several IEEE, Elsevier and Wiley journals in the areas of communications, defense and signal processing.
He has also served as reviewer and TPC member for several IEEE conferences.
\end{IEEEbiography}


\begin{IEEEbiography}[{\includegraphics[width=3in,height=1.3in,clip,keepaspectratio]{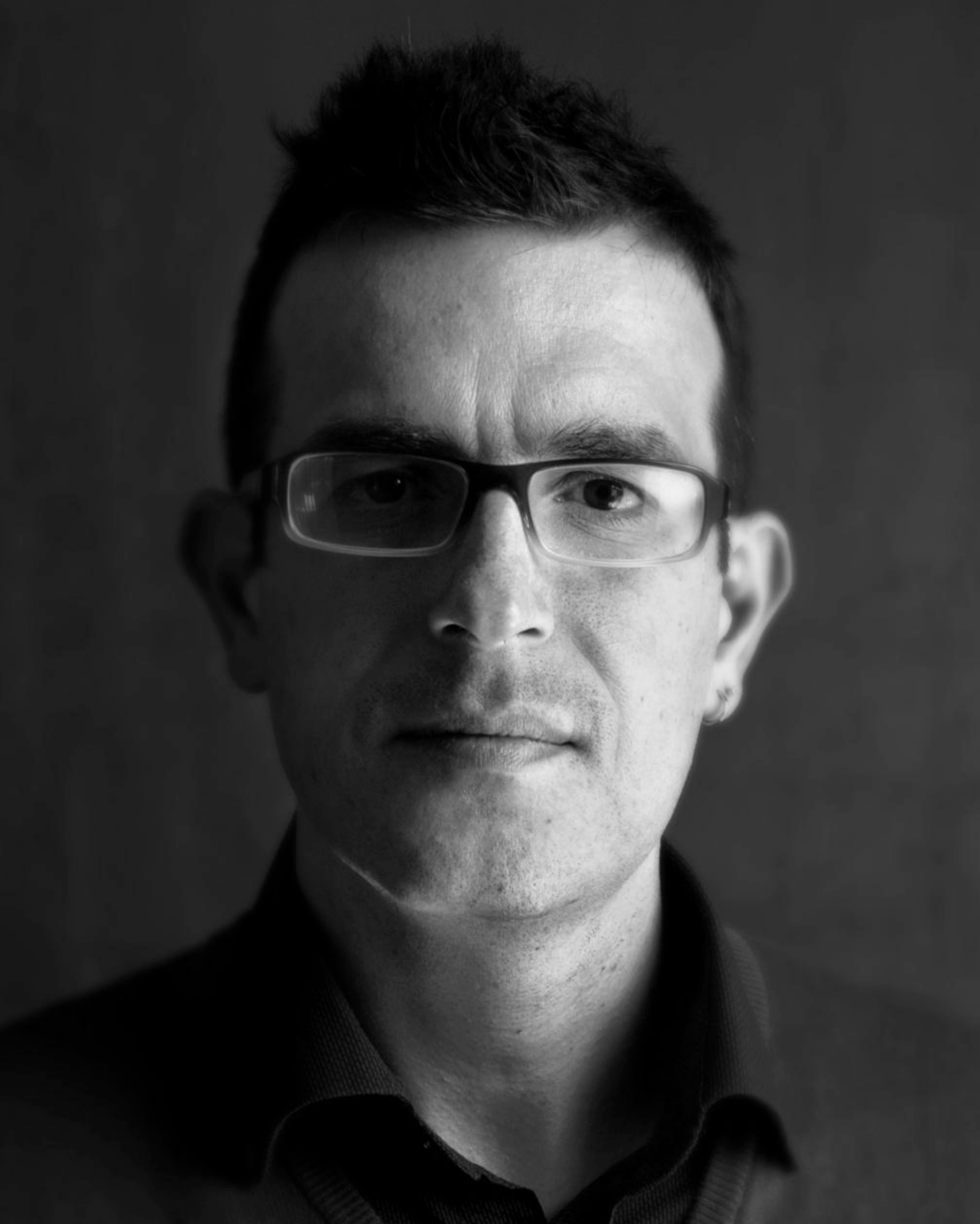}}]{Gianmarco~Romano (M'11)}
is currently Assistant Professor at the Department of Information Engineering, Second University of Naples, Aversa (CE), Italy. 
He received the ``Laurea'' degree in Electronic Engineering from the University of Naples ``Federico II'' and the Ph.D. degree from the Second University of Naples, in 2000 and 2004, respectively.
From 2000 to 2002 he has been Researcher at the National Laboratory for Multimedia Communications (C.N.I.T.) in Naples, Italy. 
In 2003 he was Visiting Scholar at the Department of Electrical and Electronic Engineering, University of Connecticut, Storrs, USA. 
Since 2005 he has been with the Department of Information Engineering, Second University of Naples and in 2006 has been appointed Assistant Professor. 
His research interests fall within the areas of signal processing and communications.
\end{IEEEbiography}

\vfill

\end{document}

%% file: Cooperative_Spectrum_Sensing---sec1.tex
\section{Introduction}
\label{sec:introduction}

\PARstart{C}{ognitive} radio is an emerging paradigm for wireless communications aiming at an efficient utilization of the spectrum \cite{mitola, haykin}.
A cognitive radio is an intelligent system for wireless communications capable to adapt to the dynamic environment for high reliability.
A crucial point is the utilization of the so-called {\em spectrum holes} or {\em white spaces}, i.e. those frequency bands assigned to a primary user (PU) and that at particular time periods are not used by that PU.

To this aim, large efforts in the recent literature have been devoted to spectrum sensing \cite{akyildiz2006, akyildiz2011, axell}, i.e. developing techniques for detecting spectrum holes in order to allow opportunistic utilization by unlicensed secondary users (SUs) without affecting the PU.
Energy detection \cite{urkowitz, digham} is one of the most popular techniques for local detection of the PU by the single SU, due to its simplicity.
Alternative detection techniques rely on ciclostationarity of communication signals \cite{lunden}.
Furthermore, collaborative spectrum sensing \cite{mishra} represents the main framework in order to mitigate (through spatial diversity) the detrimental effects due to fading, shadowing, out-of-range, etc., that the SUs may experience.

Centralized collaborative spectrum sensing \cite{unnikrishnan, quan, letaief} is one possible architecture in which a fusion center collects the individual measurements or decisions by each SU through a reporting channel and combines them to determine the presence/absence of the PU.
Decentralized alternatives have also been proposed \cite{li}.
Data fusion at the fusion center may be employed through soft combining \cite{ma} or hard combining \cite{zhang}.
In the latter case, SUs make local decisions and send a binary information to the decision fusion center (DFC).
It is worth mentioning that different approaches for spectrum sensing, exploiting results from game theory, have been also considered in the recent literature \cite{saad}.

Spatial diversity through multiple antennas at both transmit and receive locations, employing a multiple-input multiple-output (MIMO) system, is a common technique for communications in wireless environments usually affected by multipath fading \cite{goldsmith}.
User cooperation \cite{sendonaris} has been introduced as a technique providing spatial diversity through building virtual antenna arrays even in the case of single-antenna users.
Various cooperative protocols have been studied in the literature, such as amplify-and-forward and decode-and-forward \cite{laneman2003, laneman2004}, coded cooperation \cite{nosratinia2004, nosratinia2006}, signal and code superposition \cite{larsson, costello, salvorossi2007, salvorossi2010}.

It is worth noticing that in the cognitive-radio literature the terms ``collaborative'' and ``cooperative'' are used interchangeably.
Differently, in this work we use the former for the usual distributed detection framework in cognitive-radio scenarios, while the latter for the multi-access technique developed in multiuser-communication scenarios.
Integration of cooperation algorithms into collaborative spectrum sensing was also discussed in \cite{ganesan} where the focus was to increase the agility, i.e. the detection time, of the cognitive network. 
However, amplify-and-forward was considered while our paper assumes decode-and-forward.

Distributed detection was typically investigated in wireless sensor networks \cite{chen} where the common architecture is a parallel access channel (PAC), i.e. the sensors are assigned orthogonal channels for transmission.
Recently, the advantage of multiple antennas to be used over a multiple access channel (MAC) in distributed detection problems has been faced in \cite{banavar} where asymptotic techniques are considered to derive error exponents, in \cite{ciuonzo} where  decode-and-fuse vs. decode-then-fuse approaches are compared, and in \cite{ciuonzon} where energy detection properties were investigated.

Starting from the results in \cite{ciuonzo} as a general framework for distributed decision fusion, and referring to a centralized architecture for collaborative spectrum sensing, we analyze the impact of user cooperation on system performance.
More specifically, we present four different protocols for collaborative spectrum sensing that combine orthogonal/non-orthogonal (i.e. interfering/non-interfering) and cooperative/non-cooperative transmissions at transmit locations and employ MIMO decision fusion at receive location.
The first two protocols refer to the standard centralized fusion model for collaborative spectrum sensing in which each SU transmits its local decision to the DFC: (i) in the first case, we consider interfering reporting channels, i.e. an equivalent MAC from the SUs to the DFC; (ii) in the second case, we consider non-interfering reporting channels, i.e. an equivalent PAC from the SUs to the DFC.
The other two protocols are obtained through the modification of the first two by means of user cooperation between the SUs, thus giving rise to: (iii) a cooperative MAC (CMAC) and (iv) a cooperative PAC (CPAC) from the SUs to the DFC, respectively.
The protocols, denoted in the following via the equivalent reporting channel acronym, are compared in terms of complementary receiving operating characteristic (CROC) taking into account also {\em their different spectral efficiency}.
CROCs are evaluated both through Monte Carlo simulations and theoretical results.

It is worth mentioning that, similarly to our approach, the work in \cite{suzuki} also focuses on the reporting channel in a spectrum-sensing context and exploits results from decision fusion, though the system is collaborative and not cooperative.
Also, the work in \cite{peh2011} design power allocation strategies in a scenario with one single pair of SU transmitter and receiver that cooperate to improve detection of PU activity.
In our work, the cooperative frame among the SUs is built along the same lines of the protocols described in \cite{salvorossi2011}, though other cooperative protocols may be considered.

The outline of the paper is the following: 
in Sec.~\ref{sec:systemmodel} we present the system model under investigation underlining the differences of the four considered protocols; 
we derive the statistics for the decision fusion at the receiver in Sec.~\ref{sec:decisionfusion} and also describe a theoretical framework for performance analysis; 
Sec.~\ref{sec:simulationresults} highlights and compares the performance of the proposed protocols;
Sec.~\ref{sec:conclusion} gives some concluding remarks;
all proofs are found in the Appendix.

\textit{Notation --} 
Lower-case bold letters denote vectors, with $a_n$ denoting the $n$th element of $\a$; 
upper-case bold letters denote matrices, with $A_{n,m}$ denoting the $(n,m)$th element of $\A$; 
$\I_N$ denotes the $N\times N$ identity matrix;  
$\O_N$ denotes the $N\times N$ null matrix;  
$\i_N^{(n)}$ denotes the $n$th column of $\I_N$; 
$\0_{N}$ and $\1_{N}$ denote the $N$-length vectors whose elements are $0$ and $1$, respectively; 
${\rm diag}(\a)$ denotes a diagonal matrix with $\a$ on the main diagonal; 
${\rm diag}(\A_1,\ldots,\A_N)$ denotes a block diagonal matrix with matrix $\A_n$ denoting the $(n,n)$th block; 
$\Ex\{\cdot\}$, $(\cdot)^{\ast}$, $(\cdot)^{t}$ and $(\cdot)^{\dag}$ denote expectation, conjugate, transpose and conjugate transpose operators; 
$\otimes$ denotes the Kronecker matrix product; 
${\rm det}(\A)$ is the determinant of $\A$; 
$j$ is the imaginary unit; 
${\cal A}^2$ denotes the Cartesian square of the set ${\cal A}$; 
$\sim\CN(\mub,\Sigmab)$ means ``distributed according to a circular symmetric complex normal distribution with mean $\mub$ and covariance $\Sigmab$''.

%% file: Cooperative_Spectrum_Sensing---sec2.tex
\section{System Model}
\label{sec:systemmodel}

\begin{figure}[t]\centering
	\includegraphics[width=1.0\linewidth]{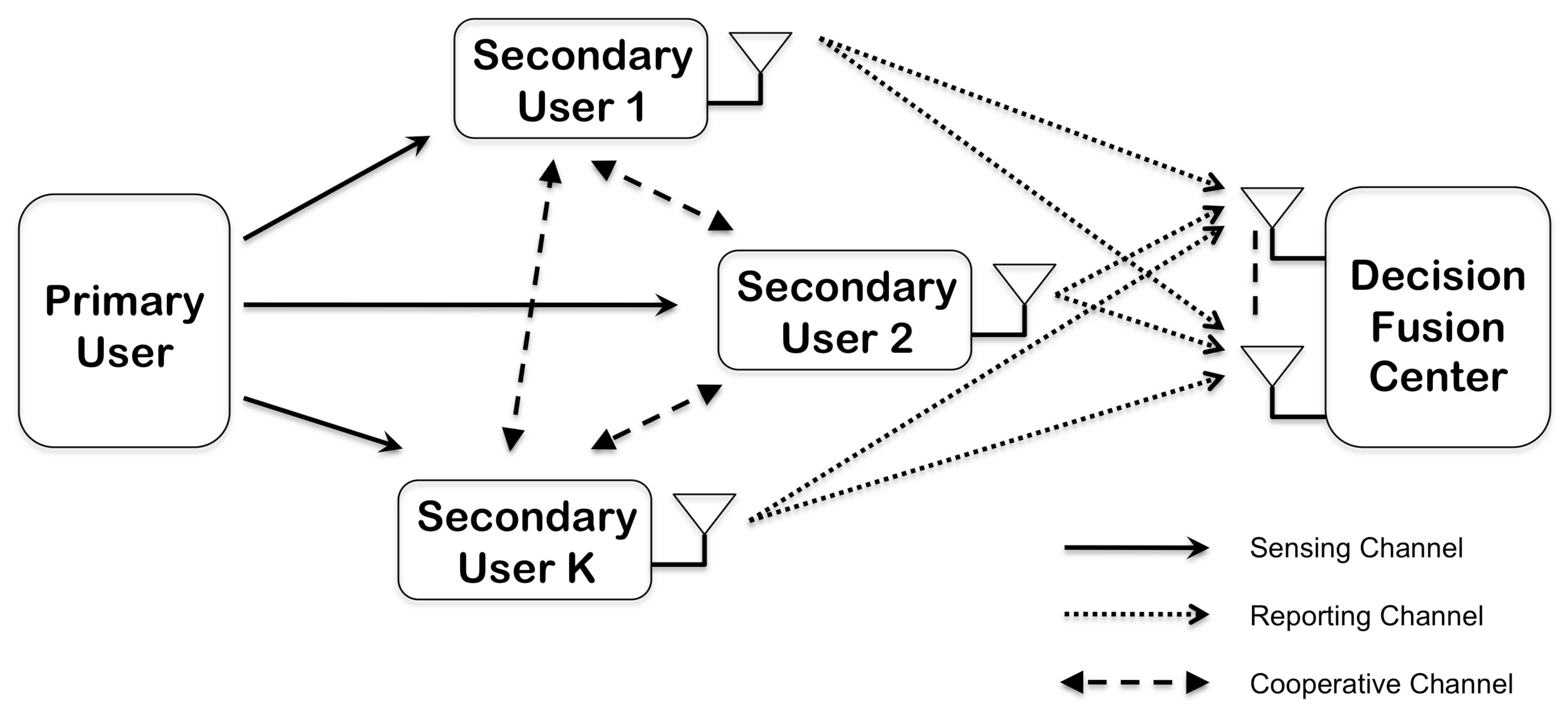}
	\caption{Cooperative spectrum sensing through MIMO decision fusion.}
	\label{fig:scenario}
\end{figure}

\begin{figure*}[t]\centering
	\includegraphics[width=1.0\linewidth]{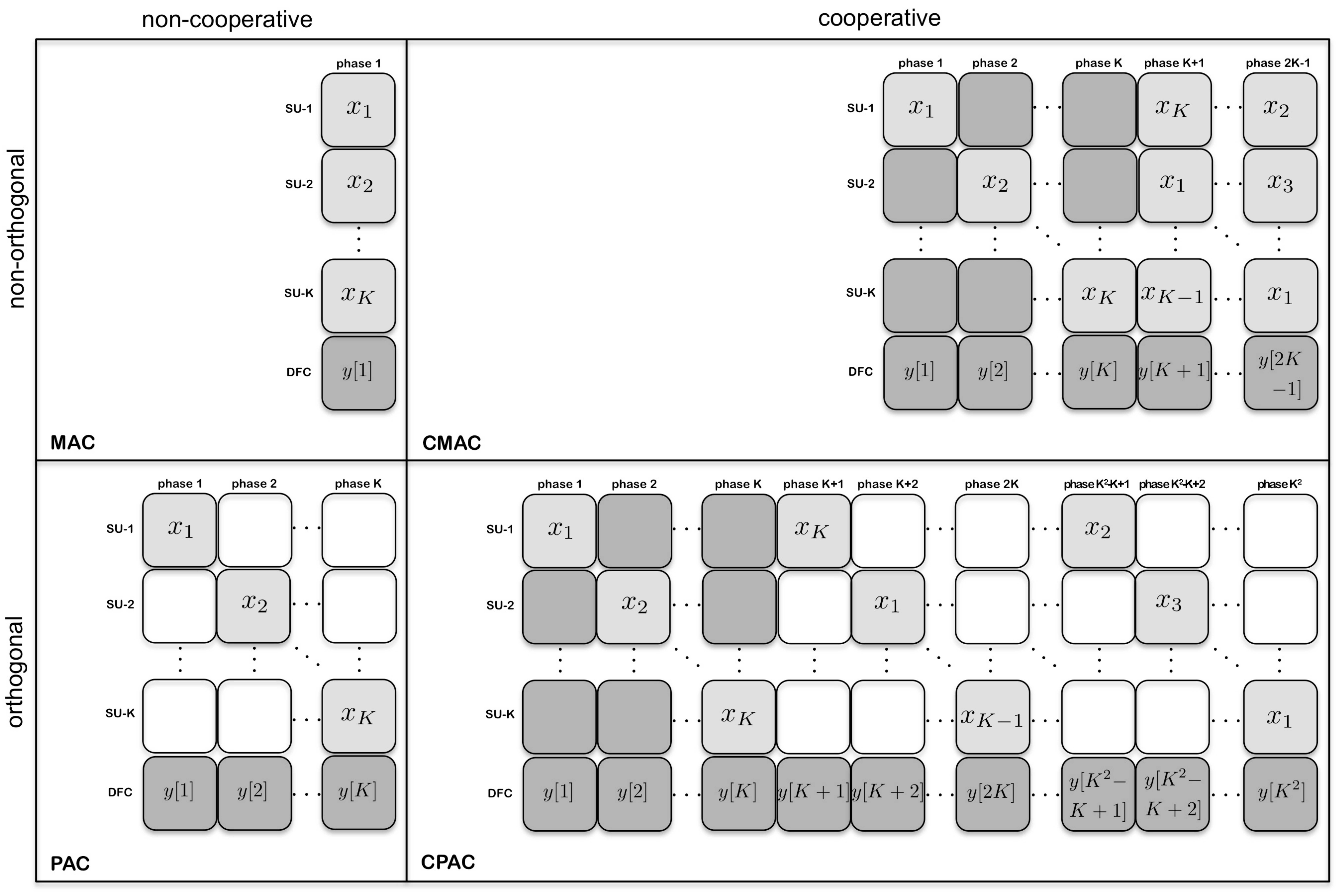}
	\caption{Structure of the communication frame for each protocol. White blocks represent inactive slots, grey blocks represent active slots (light grey for transmitting and dark grey for receiving).}
	\label{fig:coopframe}
\end{figure*}

We consider a system with $K$ (unauthorized) SUs that want to transmit in a licensed band provided that the (authorized) PU is silent.
The event that the PU is silent or active is denoted ${\cal H}_0$ or ${\cal H}_1$, and the corresponding a-priori probabilities are denoted $p_0$ and $p_1=1-p_0$, respectively. 
A DFC provides a reliable decision about the activity of the PU on the basis of the decisions taken locally (and transmitted to the DFC itself) by the SUs.
We assume that the local sensing and decision process is fully described by the local probability of detection ($p_d$) and the local probability of false alarm ($p_f$), both assumed to be stationary, identical, and conditionally independent for the group of $K$ SUs.
Each SU is equipped with one single transmit antenna, while the DFC is equipped with $N$ receive antennas.
The scenario is illustrated in Fig.~\ref{fig:scenario} where the communication process on the reporting channel may be viewed as a $K \times N$ MIMO system.

We assume half-duplex transmissions, in which users terminals cannot transmit and receive simultaneously. 
We also consider the possibility that the $K$ SUs cooperate in order to increase spatial diversity on the reporting channel.
We consider a block-fading channel, in which symbols related to the transmission of one single local decision by each SU undergo one single channel realization between a given pair of transmit/receive antennas.

We denote $\x=\left(x_1, x_2,\ldots, x_K\right)^{t}$, where $x_k\in\{-1,+1\}$ is the BPSK symbol transmitted by the $k$th SU and represents its local decision on the PU being silent/active, respectively.
Also, we denote $\h_k=\left(H_{1,k},\ldots,H_{N,k}\right)^{t}\sim\CN\left(\0_N,\I_N\right)$ where $H_{n,k}$ is the channel coefficient between the $k$th SU and the $n$th receive antenna of the DFC, and assume that $\{\h_1, \h_2, \ldots, \h_K\}$ are statistically independent.
More specifically, each considered protocol involves a different total number of transmission phases ($L$) in the generic communication frame. 
The communication frame for the general case of $K$ SUs is shown in Fig.~\ref{fig:coopframe}. 
The generic $\ell$th phase provides an $N-$length vector of received signals at the DFC, denoted $\y[\ell]=\left(y_1[\ell],\ldots,y_N[\ell]\right)^{t}$ where $y_n[\ell]$ is the signal received by the $n$th receive antenna.
Analogous notation is used for the noise contribution $\w[\ell]=\left(w_1[\ell],\ldots,w_N[\ell]\right)^{t}\sim\CN\left(\0_N,\sigma_w^2\I_N\right)$.
The discrete-time signal model (after matched filtering and sampling) for the received signal at the DFC is
\begin{align}
 	\y=\alpha\H_{eq}\x + \w \;,
	\label{eq:dtsigmod}
\end{align}
where $\y=\left(\y[1]^t,\ldots,\y[L]^t\right)^{t}$ collects the signals at the various receive antennas, $\w=\left(\w[1]^{t},\ldots,\w[L]^{t}\right)^{t}$ denotes the corresponding noise contribution, while $\alpha$, $\H_{eq}$, and $\sigma_w^2$ denote the scaling factor, the equivalent channel matrix, and the equivalent noise variance, respectively, and are expressed explicitly in the following subsections, depending on the considered protocol.
The size of the model in Eq.~(\ref{eq:dtsigmod}), i.e. the length of the vector of received signals ($\y$), is then $M=LN$.

Two criteria are considered here to determine the scaling factor ($\alpha$): (i) enforcing unitary average transmitted power per single user, in this case the ``power'' scaling factor will be denoted $\alpha_p$; (ii) enforcing unitary transmitted energy per single user, in this case the ``energy'' scaling factor will be denoted $\alpha_e$.
The first criterion is commonly used in the context of wireless communications \cite{laneman2004} or in general when the limitation of power emission, e.g. due to regulatory restrictions, is a crucial issue.
The second criterion is commonly used in the context of wireless sensor networks \cite{schober} or in general when the battery consumption is more critical than other aspects.
More specifically, we denote ${\cal P}$ and ${\cal E}$ the transmitted power and energy per user in the generic communication frame, respectively, and also $T_f$ and $T$ the time duration for the generic communication frame and the single phase, respectively.
Furthermore, we denote $q$ the number of phases in which the generic SU is transmitting (either his own information or some other partner information) in the generic communication frame ($q\leq L$).
It is apparent that 
\begin{align}
 	T_f=LT \;,\;~~ {\cal E}=q\alpha^2 \;,\;~~ {\cal P}=\frac{{\cal E}}{T_f}=\frac{q\alpha^2}{LT} \;.
\end{align}

It is worth noticing that the four considered protocols require each a different overhead for their implementation, thus exhibit different spectral efficiencies.
One possibility to compare them would be to evaluate the achievable throughput for each of them, along the same lines of \cite{peh2008, peh2009}.
Alternatively, we prefer to limit the analysis at the physical layer and take into account the impact of the overhead through scaling the noise power accordingly, as in \cite{salvorossi2011}.
The equivalent noise variance is defined as $\sigma_w^2=N_o/\eta$ where $N_o$ is the effective variance of the additive Gaussian noise and  $\eta=K/L$ bits-per-channel-use (bpcu) is the spectral efficiency of the considered protocol. 
Such a normalization, see also Section~\ref{ssec:snr}, allows fairer performance comparison, taking into account the different spectral efficiency of the different protocols.

Throughout the paper  we will consider that cooperation is employed through decode-and-forward assuming perfect communications between the SUs which is a realistic situation since cooperation usually takes place between terminals that are separated by a reliable channel \cite{salvorossi2011}.
Finally, we assume perfect frame and phase synchronization between the SUs and perfect channel state information at the DFC, as in \cite{salvorossi2011}.

\subsection{MAC protocol}
The protocol involves $L=1$ phase and each SU transmits during $q=1$ phase, thus the vector of received signals is $\y=\y[1]$, the size of the model is $M=N$, and the spectral efficiency of the protocol is $\eta=K$ (bpcu).
The power and energy scaling factors, and the equivalent noise variance are
\begin{align}
 	\alpha_p=\sqrt{T} \;,\;~~ \alpha_e=1 \;,\;~~ \sigma_w^2=\frac{1}{K}N_o \;,
\end{align}
while the equivalent channel matrix is
\begin{align}
	\H_{eq}=\left(\begin{array}{cccc} 
	\h_1 & \h_2 & \cdots & \h_K 
	\end{array}\right) \;.
	\label{eq:mac_set}
\end{align}
The structure of the communication frame is shown in the upper-left quadrant of Fig.~\ref{fig:coopframe}.
Both SUs transmit simultaneously and do not act as a relay for each other: the protocol is {\em non-orthogonal} and {\em non-cooperative}.

\subsection{PAC protocol}
The protocol involves $L=K$ phases and each SU transmits during $q=1$ phase, thus the vector of received signals is $\y=\left(\y[1]^t,\ldots,\y[K]^t\right)^t$, the size of the model is $M=KN$, and the spectral efficiency of the protocol is $\eta=1$ (bpcu).
The power and energy scaling factors, and the equivalent noise variance are
\begin{align}
 	\alpha_p=\sqrt{KT} \;,\;~~ \alpha_e=1 \;,\;~~ \sigma_w^2=N_o \;,
\end{align}
while the equivalent channel matrix is
\begin{align}
	\H_{eq}=\left(\begin{array}{cccc} 
	\h_1 & \0_N & \cdots & \0_N \\ \0_N & \h_2 & \cdots & \0_N \\ \vdots & \vdots & \ddots & \vdots \\ \0_N & \0_N & \cdots & \h_K 
	\end{array}\right) \;.
	\label{eq:pac_set}
\end{align}
The structure of the communication frame is shown in the lower-left quadrant of Fig.~\ref{fig:coopframe}.
Each SU is silent when the other is transmitting and does not act as a relay for the other: the protocol is {\em orthogonal} and {\em non-cooperative}.

\subsection{CMAC protocol}
The protocol involves $L=2K-1$ phases and each SU transmits during $q=K$ phases, thus the vector of received signals is $\y=\left(\y[1]^t,\ldots,\y[2K-1]^t\right)^t$, the size of the model is $M=(2K-1)N$, and the spectral efficiency of the protocol is $\eta=K/(2K-1)$ (bpcu).
The power and energy scaling factors, and the equivalent noise variance are
\begin{align}
 	\alpha_p=\sqrt{\frac{2K-1}{K}T} \;,\;~~ \alpha_e=\sqrt{\frac{1}{K}} \;,\;~~ \sigma_w^2=\frac{2K-1}{K}N_o \;,
\end{align}
while the equivalent channel matrix is
\begin{align}
	\H_{eq}=\left(\begin{array}{cccc} 
	\h_1 & \0_N & \cdots & \0_N \\ \0_N & \h_2 & \cdots & \0_N \\ \vdots & \vdots & \ddots & \vdots \\ \0_N & \0_N & \cdots & \h_K \\ 
	\h_2 & \h_3 & \cdots & \h_1 \\ \h_3 & \h_4 & \cdots & \h_2 \\ \vdots & \vdots & \ddots & \vdots \\ \h_K & \h_1 & \cdots & \h_{K-1} \end{array}\right) \;.
	\label{eq:cmac_set}
\end{align}

The structure of the communication frame is shown in the upper-right quadrant of Fig.~\ref{fig:coopframe}.
Each SU is silent when the other is transmitting its own source information and both act as a relay for each other employing interfering transmissions: the protocol is {\em non-orthogonal} and {\em cooperative}.

\subsection{CPAC protocol}
The protocol involves $L=K^2$ phases and each SU transmits during $q=K$ phases, thus the vector of received signals is $\y=\left(\y[1]^t,\ldots,\y[K^2]^t\right)^t$, the size of the model is $M=K^2N$, and the spectral efficiency of the protocol is $\eta=1/K$ (bpcu).
The power and energy scaling factors, and the equivalent noise variance are
\begin{align}
 	\alpha_p=\sqrt{KT} \;,\;~~ \alpha_e=\sqrt{\frac{1}{K}} \;,\;~~ \sigma_w^2=KN_o \;,
\end{align}
while the equivalent channel matrix is
\begin{align}
	\H_{eq}=\left(\begin{array}{cccc} 
	\h_1 & \0_N & \cdots & \0_N \\ \0_N & \h_2 & \cdots & \0_N \\ \vdots & \vdots & \ddots & \vdots \\ \0_N & \0_N & \cdots & \h_K \\
	\h_2 & \0_N & \cdots & \0_N \\ \0_N & \h_3 & \cdots & \0_N \\ \vdots & \vdots & \ddots & \vdots \\ \0_N & \0_N & \cdots & \h_1 \\
	\vdots & \vdots & \ddots & \vdots \\ 
	\h_K & \0_N & \cdots & \0_N \\ \0_N & \h_1 & \cdots & \0_N \\ \vdots & \vdots & \ddots & \vdots \\ \0_N & \0_N & \cdots & \h_{K-1}  
	\end{array}\right) \;.
	\label{eq:cpac_set}
\end{align}
The structure of the communication frame is shown in the lower-right quadrant of Fig.~\ref{fig:coopframe}.
Each SU is silent both when the other is transmitting its own source information and when acting as a relay: the protocol is {\em orthogonal} and {\em cooperative}.

\subsection{Signal-to-noise ratio (SNR) on the reporting channel}
\label{ssec:snr}
When comparing the performance of the four protocols in various system configurations, we will refer to two different definitions for the signal-to-noise ratio (SNR).
In the case of power constraint, the SNR is defined as the ratio between the total power transmitted in a frame ($K{\cal P}$) and the effective noise power ($N_o/T$), then SNR$_p=KT/N_o$ in agreement with \cite{laneman2003, laneman2004}.
In the case of energy constraint, the SNR is defined as the ratio between the total energy transmitted in a frame ($K{\cal E}$) and the effective noise variance ($N_o$), then SNR$_e=K/N_o$ in agreement with \cite{schober}.

Also, it is worth highlighting that the results for a given SNR have been simulated through an {\em equivalent SNR} in order to compensate for the different spectral efficiency of each protocol.
The equivalent SNR is defined replacing the effective noise variance with the equivalent noise variance in the corresponding expression, thus ${\rm SNR}_{eq}=\eta\cdot {\rm SNR}$.

%% file: Cooperative_Spectrum_Sensing---sec3.tex
\section{Decision Fusion and Performance Analysis}
\label{sec:decisionfusion}

The log-likelihood ratio of the received signal under the two hypotheses is the optimum statistic, however various approximations may be considered as valid alternatives.
We assume that maximum ratio combining (MRC) fusion rule is employed at the DFC.
MRC rule considers the simplifying assumption that local decision are perfect, i.e. $\x\in\{-\1_K,+\1_K\}$, and is based on the following test 
\begin{align}
 	\Lambda=\Re\left(\1_K^t\H_{eq}^\dag\y\right) 
	\begin{array}{c}
	{\scriptstyle \hat{{\cal H}}={\cal H}_{1}}\\\gtrless\\{\scriptstyle \hat{{\cal H}}={\cal H}_{0}}
	\end{array}
	\gamma \;,
	\label{eq:mrc}
\end{align}
where $\hat{{\cal H}}$ denotes the estimated hypothesis and $\gamma$ is a threshold. 
It is worth noticing that the assumption of perfect local decisions is used {\em only} for system design purposes, and does {\em not} mean that the system is working under such ideal conditions, thus the rule is suboptimal.
MRC rule is appealing due to numerical stability, low computational complexity, and optimum performance at low SNR \cite{ciuonzo}.
Also, MRC rule provides closed-form expressions for performance analysis \cite{ciuonzo2} which easily allows system design.
A comparison among different suboptimal fusion rules in the general context of binary decision fusion is found in \cite{ciuonzo}.

The performance of the system are evaluated in terms of probability of false alarm ($q_f$) and probability of detection ($q_d$) or probability of missed detection ($q_m=1-q_d$), defined as follows
\begin{align}
 	q_f=\Pr\left(\Lambda>\gamma|{\cal H}_0\right) \;,\;
	q_d=\Pr\left(\Lambda>\gamma|{\cal H}_1\right) \;,	
\end{align}
The threshold ($\gamma$) is typically chosen according to Bayes or Neyman-Pearson criteria \cite{kay}, i.e. minimizing the probability of error ($p_e=p_0q_f+p_1q_m$) or keeping a fixed probability of false alarm ($q_f$), respectively.

Analogously to the procedure in \cite{schober, ciuonzo2}, we compute $q_f$ through the following integral
\begin{align}
	q_f=\frac{1}{2\pi j}\int_{c-j\infty}^{c+j\infty} {\frac{1}{s}\Phi_{-\Lambda}(s|{\cal H}_0)\exp(-\gamma s)ds} \;, \label{eq:integral}
\end{align}
where $c$ is a positive constant in the region of convergence of the integral and $\Phi_{-\Lambda}(s|{\cal H}_0)$ is the moment generating function (MGF) of $\Lambda|{\cal H}_0$, i.e. the Laplace transform of the probability density function of $-\Lambda|{\cal H}_0$.
Replacing $\Phi_{-\Lambda}(s|{\cal H}_0)$ with $\Phi_{-\Lambda}(s|{\cal H}_1)$, i.e. the MGF of $\Lambda|{\cal H}_1$, within the integral in Eq.~(\ref{eq:integral}) provides $q_d$.
If closed-form expressions for both MGFs are available, the integrals are easily computed numerically through the Gauss-Chebyshev quadrature rules \cite{biglieri, annamalai}.

The MGFs of $\Lambda|{\cal H}_0$ for the MAC and PAC protocols, for arbitrary number of SUs, are shown in Eqs.~(\ref{eq:mgf_mac_0_K}) and (\ref{eq:mgf_pac_0_K}), respectively, at the top of the next page.
Also, at the top of the next page, Eqs.~(\ref{eq:mgf_cmac_0_k2}) and (\ref{eq:mgf_cmac_0_k3}) represent the MGFs of $\Lambda|{\cal H}_0$ for the CMAC protocol, in the case with $K=2$ and $K=3$ SUs, respectively.
Finally, the MGF of $\Lambda|{\cal H}_0$ for the CPAC protocol, for arbitrary number of SUs, is shown in Eq.~(\ref{eq:mgf_cpac_0_K}) at the top of the next page.
The MGFs of $\Lambda|{\cal H}_1$ are obtained replacing $p_f$ with $p_d$ in the corresponding expressions.
The derivation is shown in the Appendix.

\begin{figure*}[t]\centering\normalsize
\begin{align}	
	\Phi_{-\Lambda}(s|{\cal H}_0)=& \sum_{\nu=0}^{K} \left(\frac{1}{1+(K-2\nu)s+\left(\nu^2-K\nu-\frac{K\sigma_w^2}{4}\right)s^2}\right)^N \binom{K}{\nu}(1-p_f)^{K-\nu}p_f^\nu \label{eq:mgf_mac_0_K}\;,\\
	\Phi_{-\Lambda}(s|{\cal H}_0)=&\left(\left(\frac{1}{1+s-\frac{\sigma_w^2}{4}s^2}\right)^N(1-p_f)+\left(\frac{1}{1-s-\frac{\sigma_w^2}{4}s^2}\right)^Np_f\right)^K \label{eq:mgf_pac_0_K}\;,
\end{align}
\hrulefill
\end{figure*}
\begin{figure*}[t]\centering\normalsize
\begin{align}	
	\Phi_{-\Lambda}(s|{\cal H}_0)=&\left(\frac{1}{1+4s+(3-\sigma_w^2)s^2-\frac{3}{2}\sigma_w^2s^3+\frac{3}{16}\sigma_w^4s^4}\right)^{N}(1-p_f)^2 +\left(\frac{1}{1-\sigma_w^2s^2+\frac{3}{16}\sigma_w^4s^4}\right)^{N}2p_f(1-p_f) \nonumber\\
	&+\left(\frac{1}{1-4s+(3-\sigma_w^2)s^2+\frac{3}{2}\sigma_w^2s^3+\frac{3}{16}\sigma_w^4s^4}\right)^{N}p_f^2 \label{eq:mgf_cmac_0_k2}\;,\\
	\Phi_{-\Lambda}(s|{\cal H}_0)=&\left(\frac{1}{1+9s+(15-\frac{9}{4}\sigma_w^2)s^2+(7-\frac{15}{2}\sigma_w^2)s^3+\frac{\sigma_w^4}{4}(\frac{15}{4}\sigma_w^2-21)s^4+\frac{21}{16}\sigma_w^4s^5-\frac{7}{64}\sigma_w^6s^6}\right)^{N}(1-p_f)^3 \nonumber\\
	&+\left(\frac{1}{1+3s+(1-\frac{9}{4}\sigma_w^2)s^2-(1+\frac{5}{2}\sigma_w^2)s^3+\frac{\sigma_w^4}{4}(\frac{15}{4}\sigma_w^2+1)s^4+\frac{7}{16}\sigma_w^4s^5-\frac{7}{64}\sigma_w^6s^6}\right)^{N}3p_f(1-p_f)^2 \nonumber\\
	&+\left(\frac{1}{1-3s+(1-\frac{9}{4}\sigma_w^2)s^2+(1+\frac{5}{2}\sigma_w^2)s^3+\frac{\sigma_w^4}{4}(\frac{15}{4}\sigma_w^2+1)s^4-\frac{7}{16}\sigma_w^4s^5-\frac{7}{64}\sigma_w^6s^6}\right)^{N}3p_f^2(1-p_f) \nonumber\\
	&+\left(\frac{1}{1-9s+(15-\frac{9}{4}\sigma_w^2)s^2-(7-\frac{15}{2}\sigma_w^2)s^3+\frac{\sigma_w^4}{4}(\frac{15}{4}\sigma_w^2-21)s^4-\frac{21}{16}\sigma_w^4s^5-\frac{7}{64}\sigma_w^6s^6}\right)^{N}p_f^3 \label{eq:mgf_cmac_0_k3}\;,
\end{align}
\hrulefill
\end{figure*}
\begin{figure*}[t]\centering\normalsize
\begin{align}	
	\Phi_{-\Lambda}(s|{\cal H}_0)=& \sum_{\nu=0}^{K} \left(\frac{1}{1+(K-2\nu)s-\frac{K\sigma_w^2}{4}s^2}\right)^{KN} \binom{K}{\nu}(1-p_f)^{K-\nu}p_f^\nu \label{eq:mgf_cpac_0_K}\;,
\end{align}
\hrulefill
\end{figure*}

%% file: Cooperative_Spectrum_Sensing---sec4.tex
\section{Numerical Results}
\label{sec:simulationresults}

\begin{figure*}[t]\centering
	\subfigure[Power constraint and $K=2$ SUs.]{\includegraphics[width=0.49\linewidth]{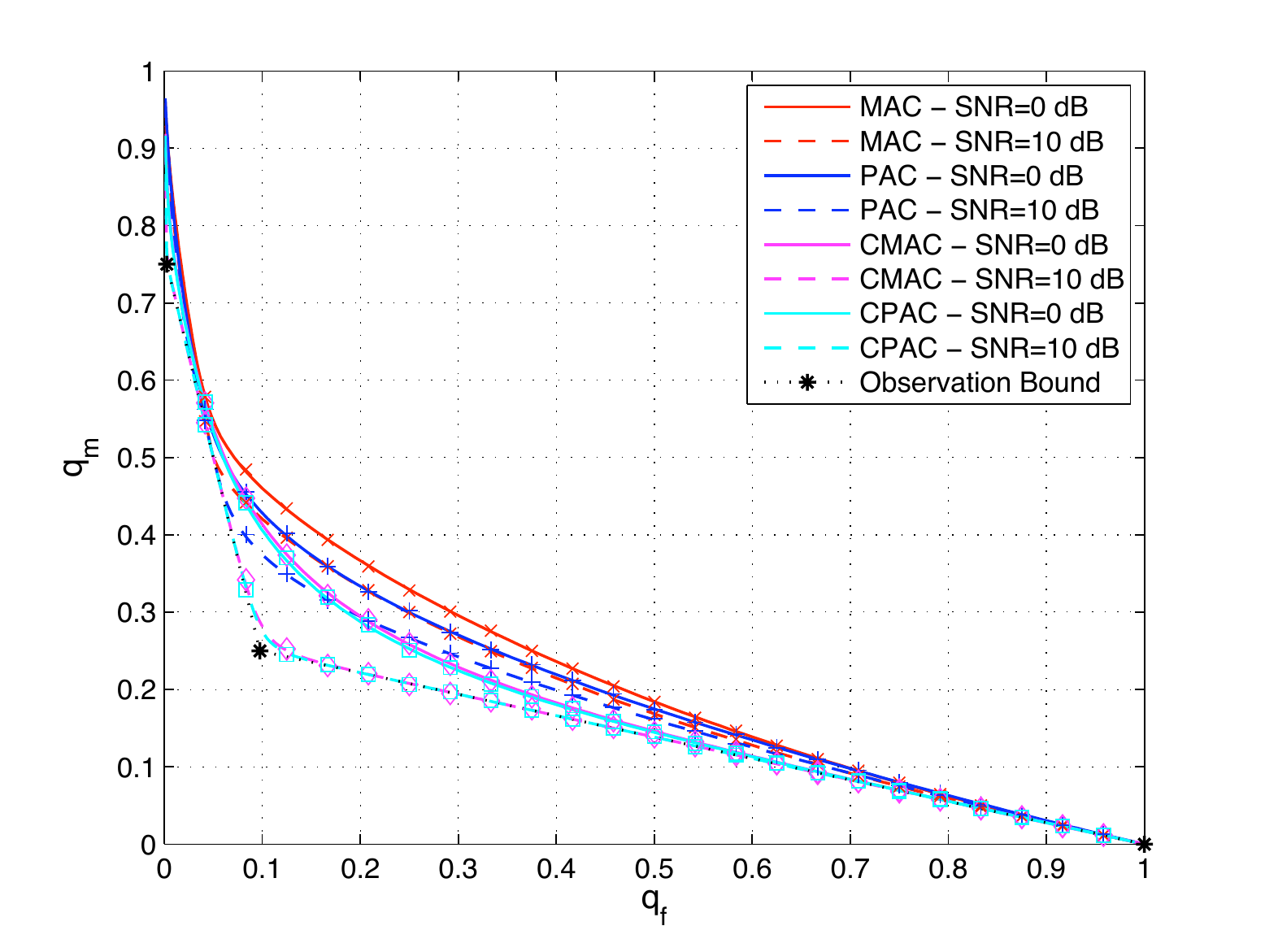}\label{fig:pCROC_N2_K2}}
	\hfill
	\subfigure[Energy constraint and $K=2$ SUs.]{\includegraphics[width=0.49\linewidth]{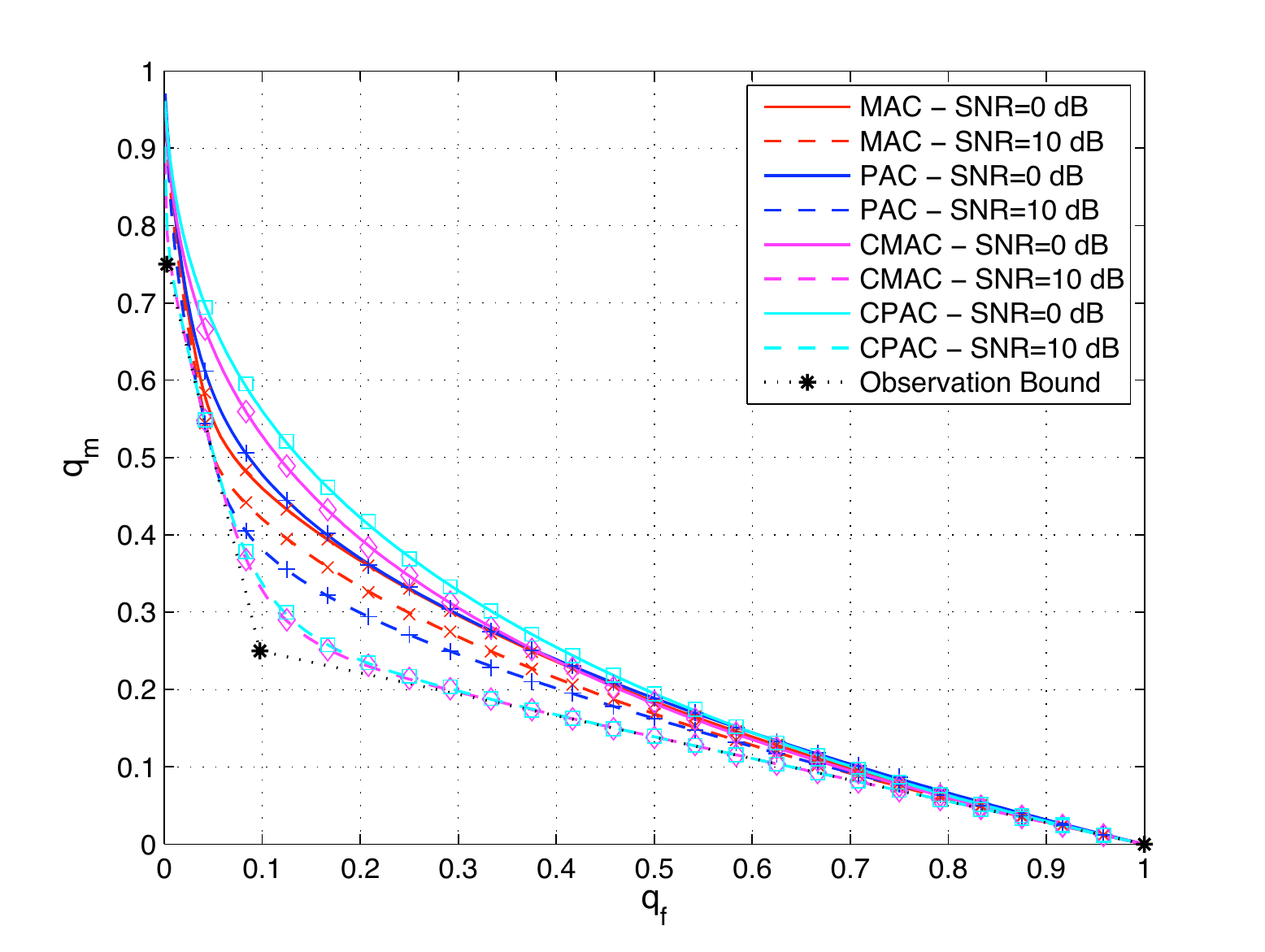}\label{fig:eCROC_N2_K2}}
	\\
	\subfigure[Power constraint and $K=3$ SUs.]{\includegraphics[width=0.49\linewidth]{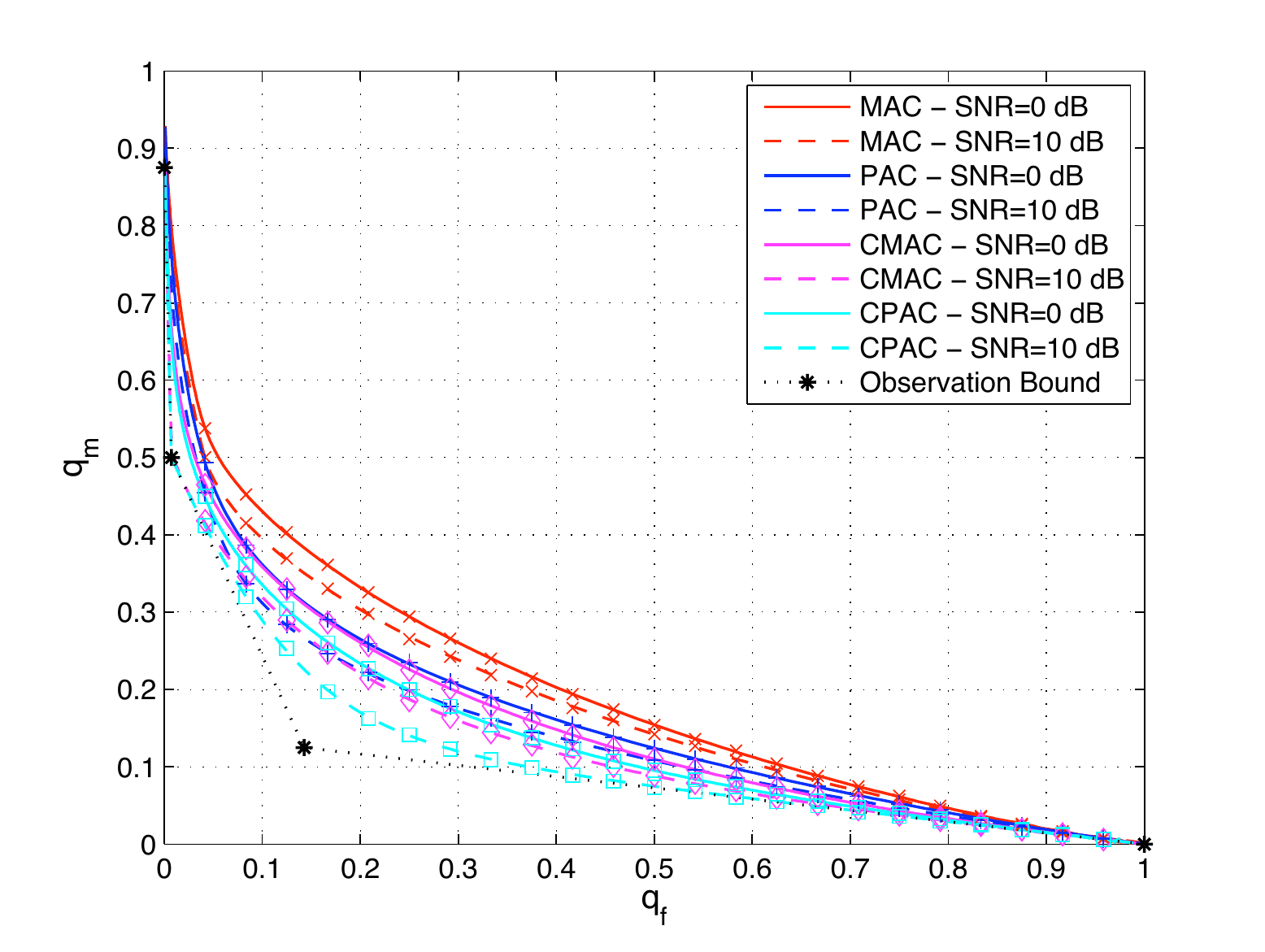}\label{fig:pCROC_N2_K3}}
	\hfill
	\subfigure[Energy constraint and $K=3$ SUs.]{\includegraphics[width=0.49\linewidth]{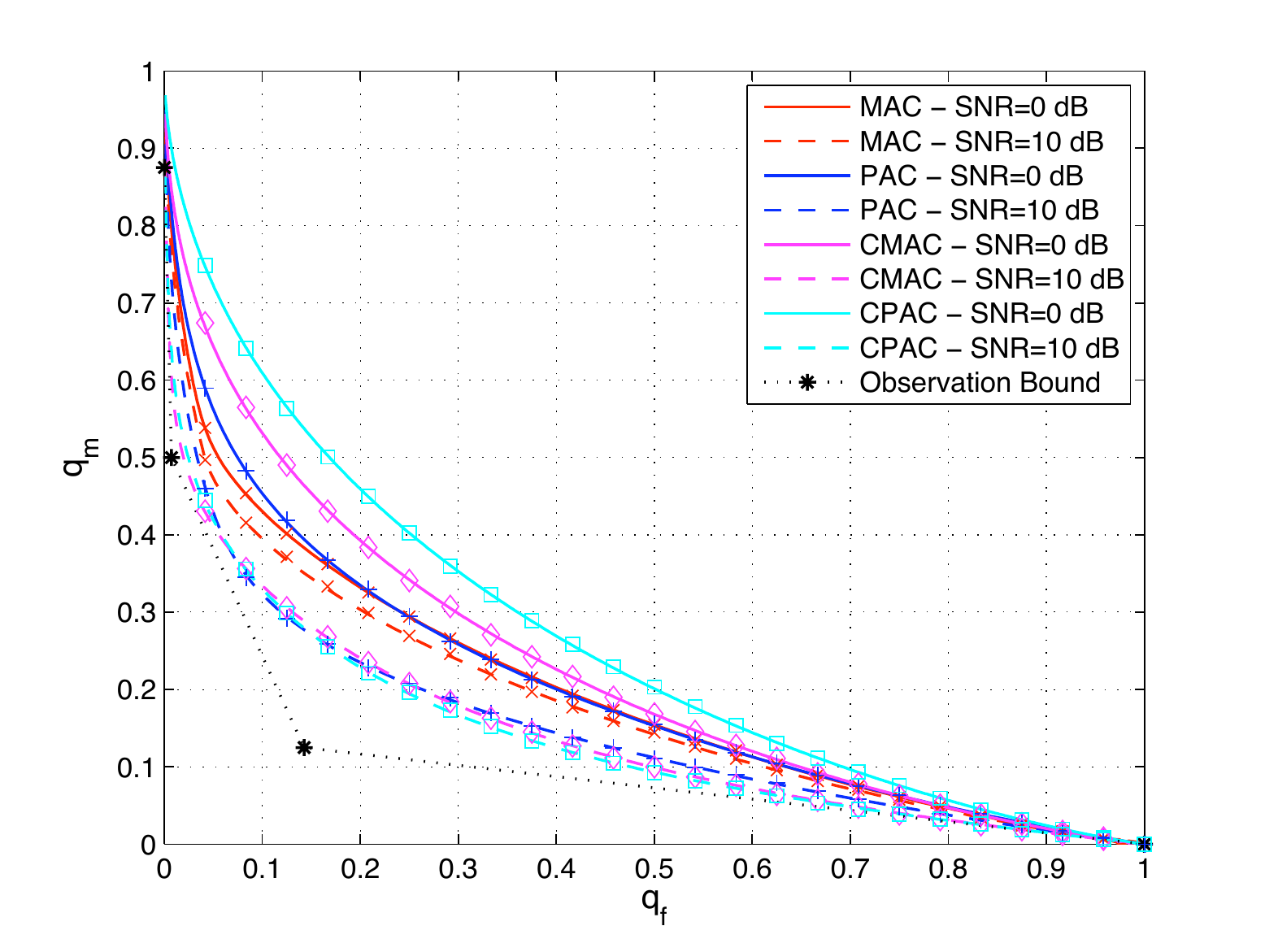}\label{fig:eCROC_N2_K3}}
	\caption{CROC for the four protocols in the case of $N=2$ receive antennas at the DFC: impact on the SNR. Lines (red for MAC, blue for PAC, magenta for CMAC, and cyan for CPAC) and markers ($\times$ for MAC, $+$ for PAC, $\Diamond$ for CMAC, and $\square$ for CPAC) refer to analytical and simulation results, respectively.}\label{fig:CROC_N2}
\end{figure*}
\begin{figure*}[t]\centering
	\subfigure[Power constraint and $K=2$ SUs.]{\includegraphics[width=0.49\linewidth]{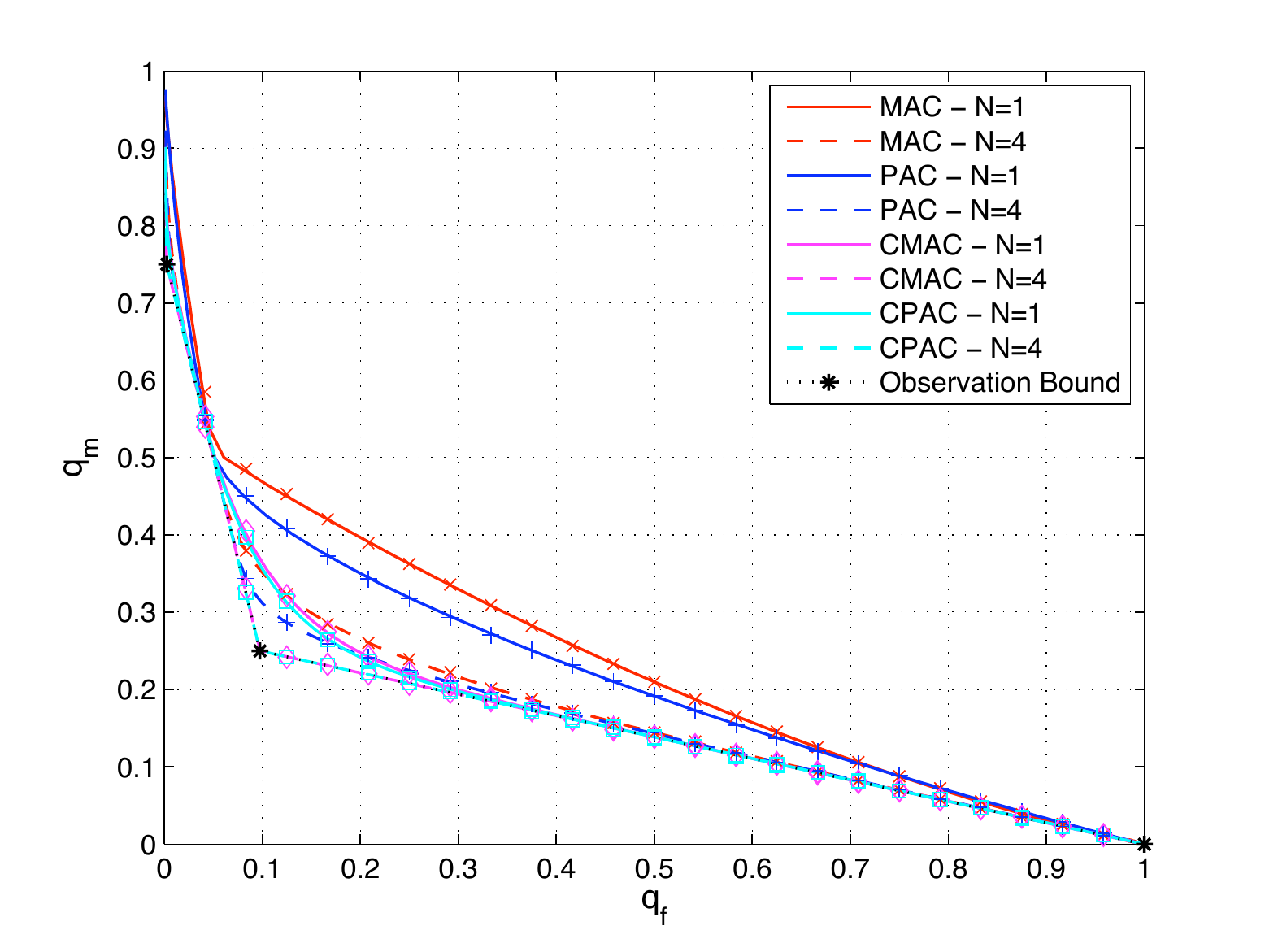}\label{fig:pCROC_10dB_K2}}
	\hfill
	\subfigure[Energy constraint and $K=2$ SUs.]{\includegraphics[width=0.49\linewidth]{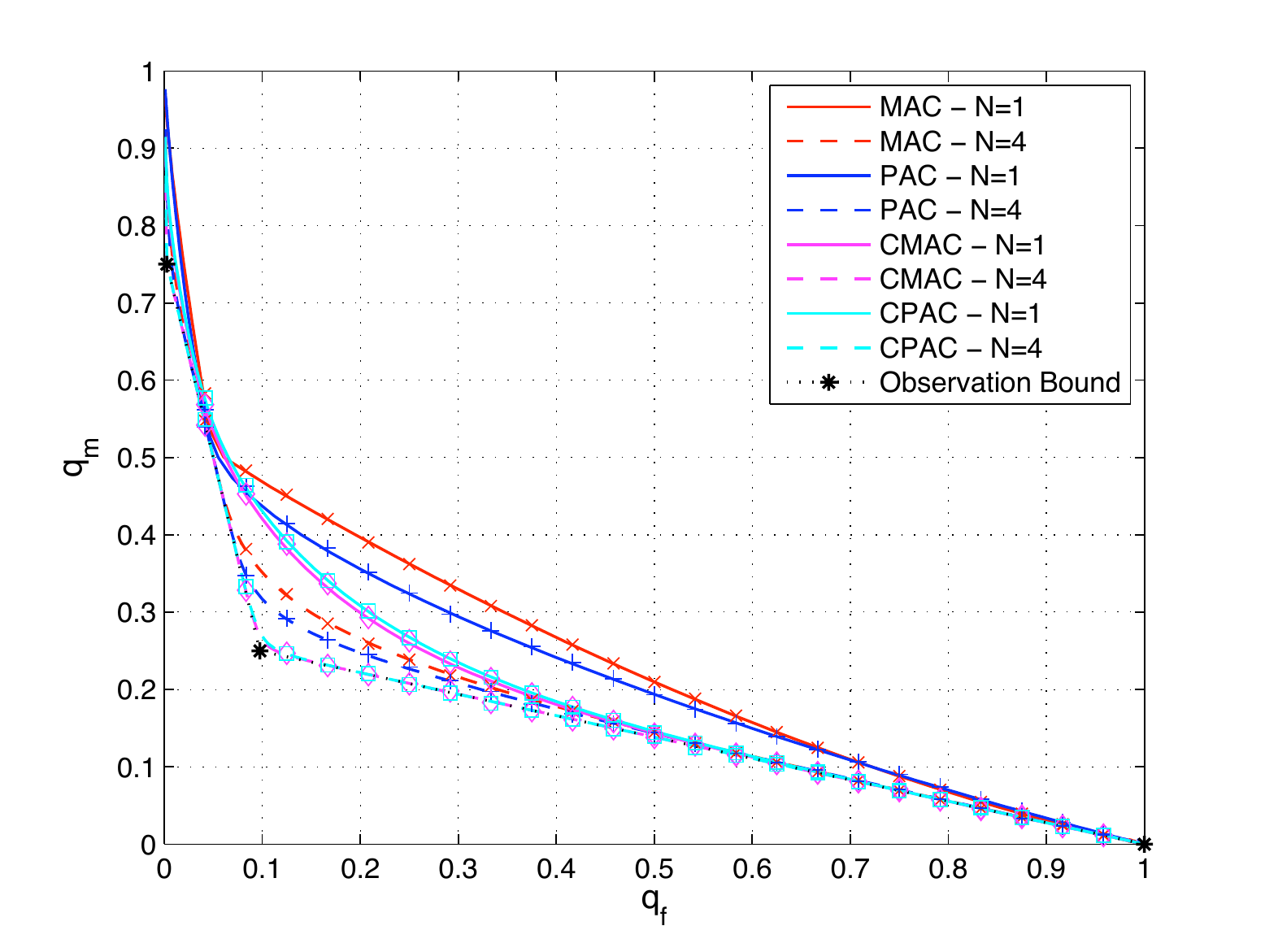}\label{fig:eCROC_10dB_K2}}
	\\
	\subfigure[Power constraint and $K=3$ SUs.]{\includegraphics[width=0.49\linewidth]{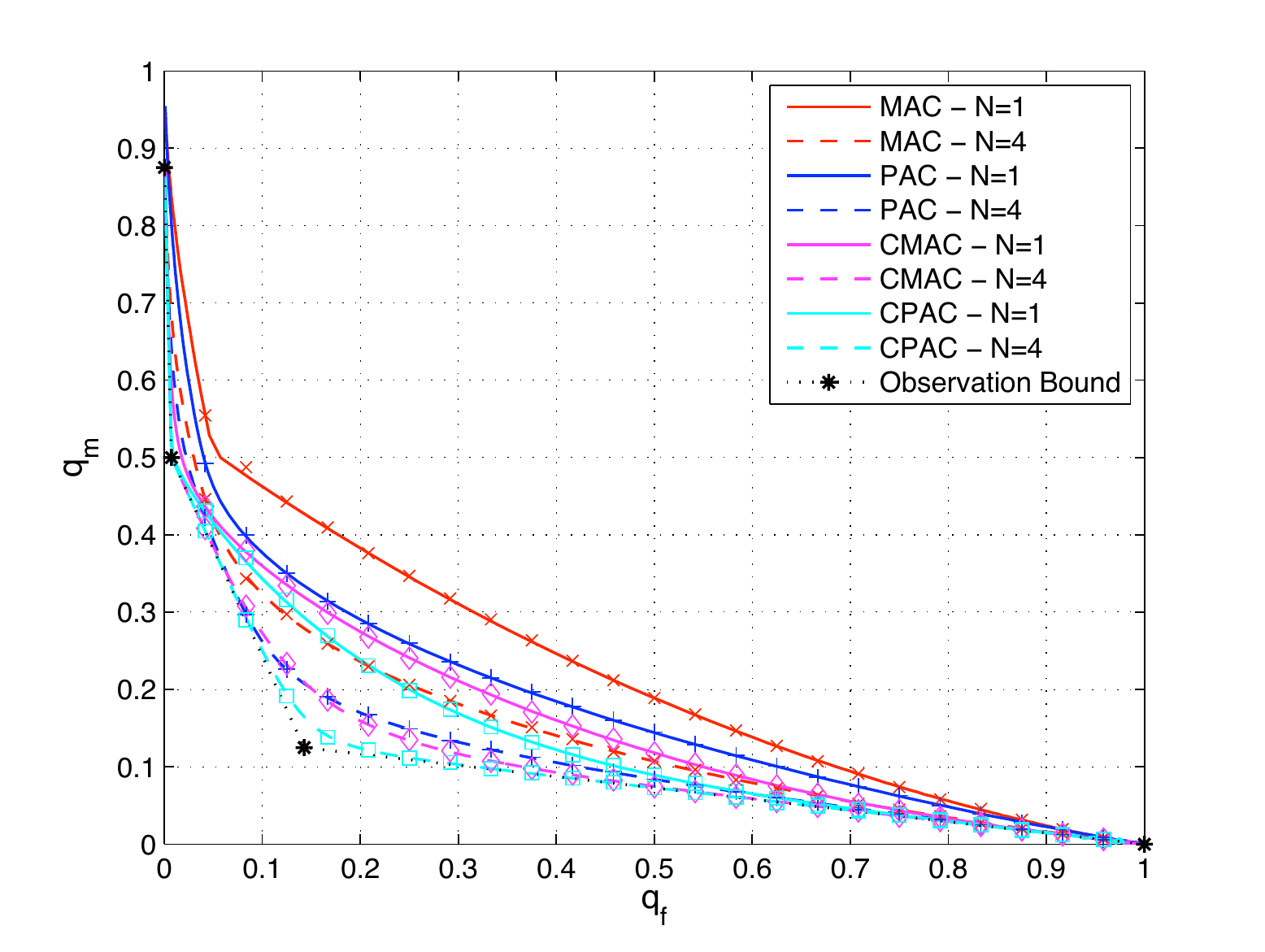}\label{fig:pCROC_10dB_K3}}
	\hfill
	\subfigure[Energy constraint and $K=3$ SUs.]{\includegraphics[width=0.49\linewidth]{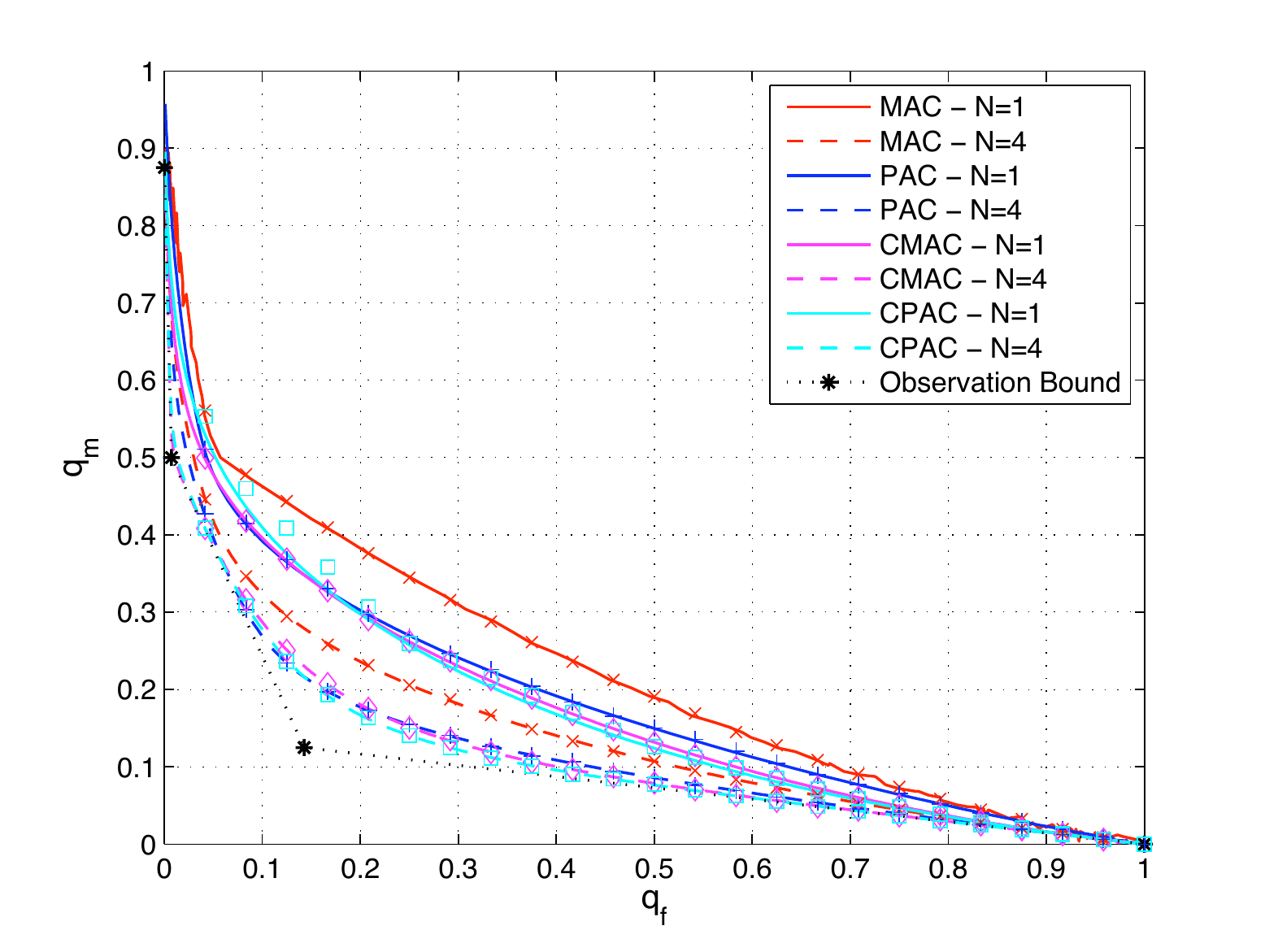}\label{fig:eCROC_10dB_K3}}
	\caption{CROC for the four protocols in the case of SNR$=10$ dB: impact on $N$. Lines (red for MAC, blue for PAC, magenta for CMAC, and cyan for CPAC) and markers ($\times$ for MAC, $+$ for PAC, $\Diamond$ for CMAC, and $\square$ for CPAC) refer to analytical and simulation results, respectively.}\label{fig:CROC_10dB}
\end{figure*}
\begin{figure*}[t]\centering
	\subfigure[MAC and CMAC comparison with power constraint.]{\includegraphics[width=0.49\linewidth]{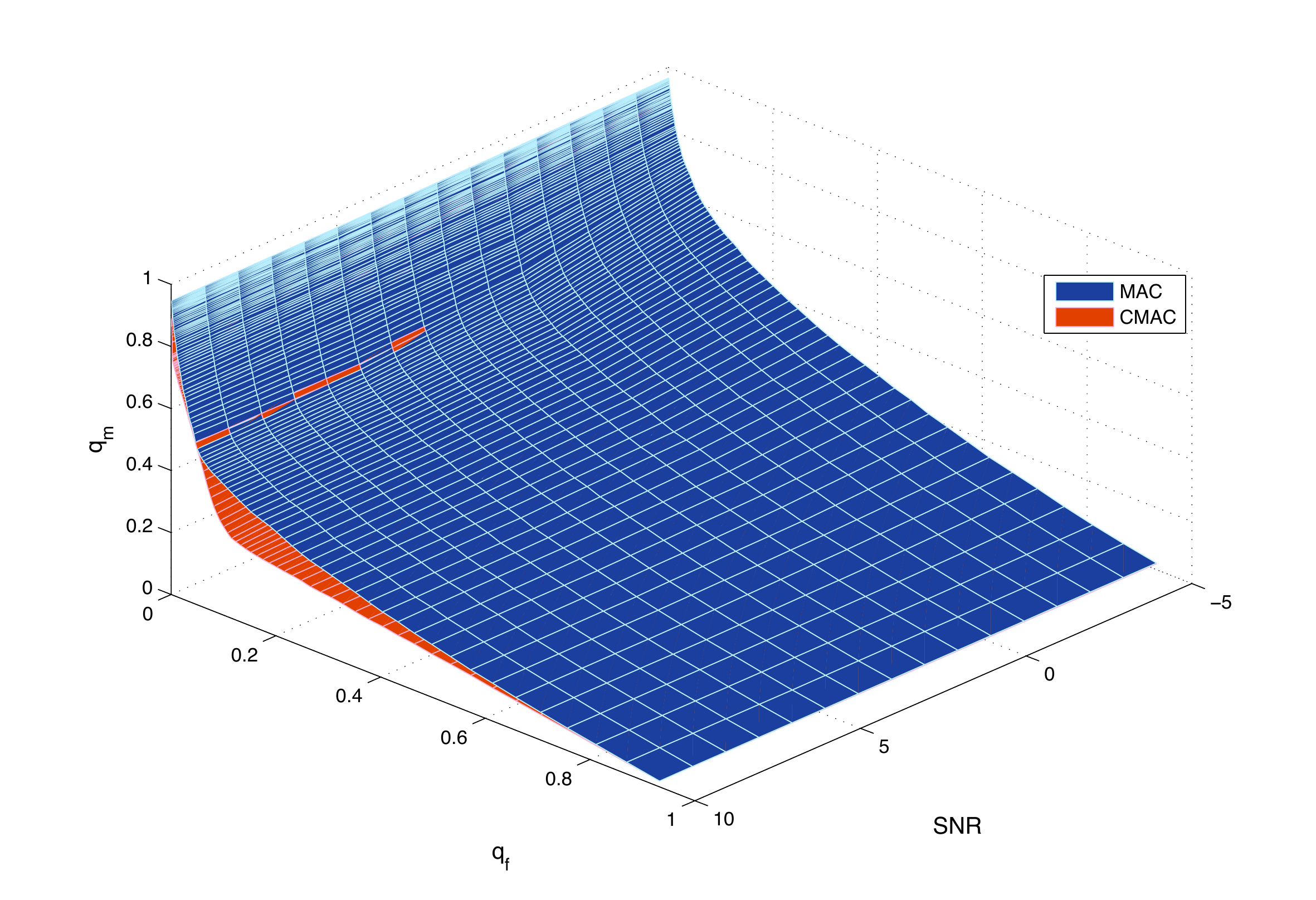}\label{fig:pMAC_pCMAC_N2}}
	\hfill
	\subfigure[MAC and CMAC comparison with energy constraint.]{\includegraphics[width=0.49\linewidth]{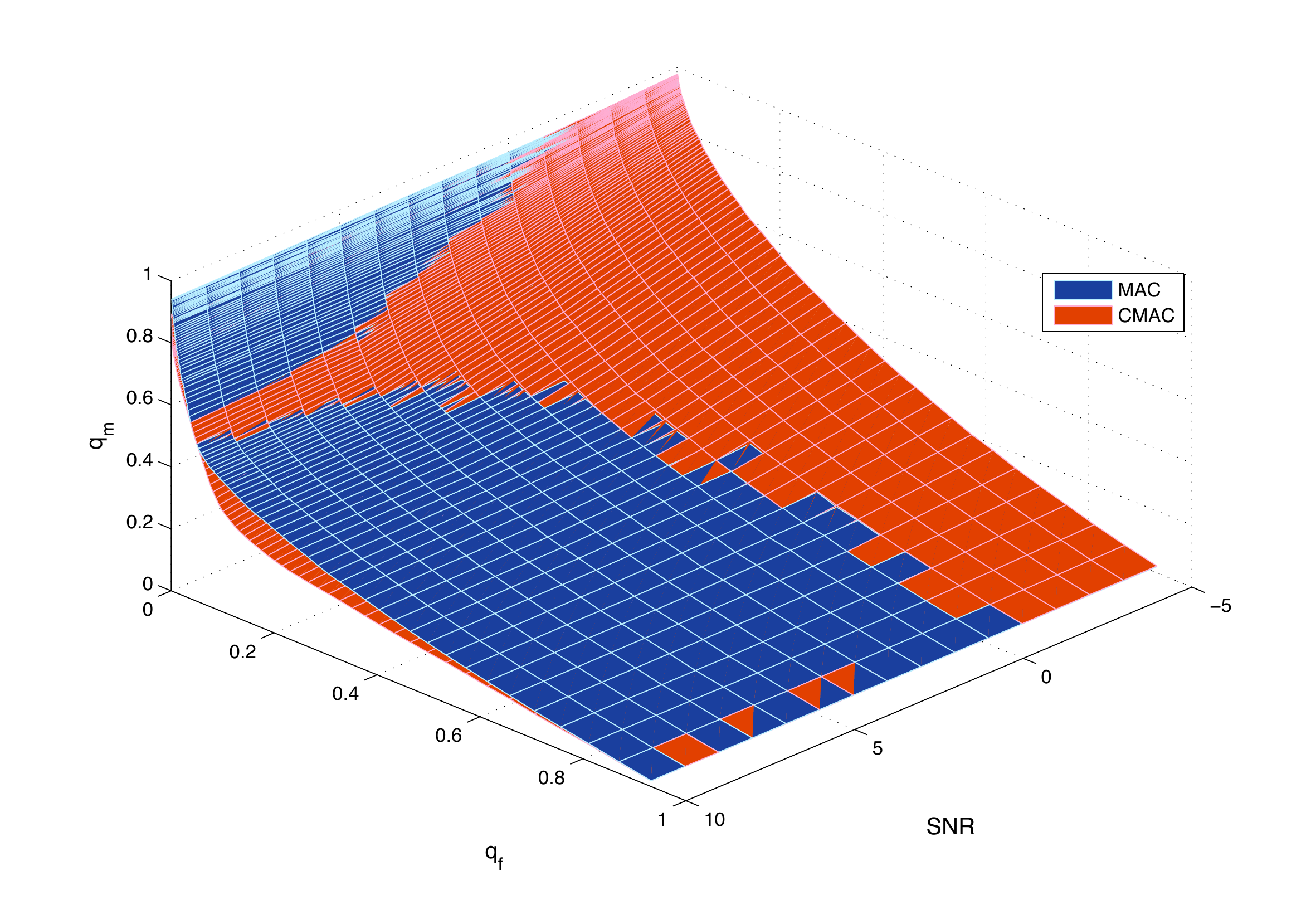}\label{fig:eMAC_eCMAC_N2}}
	\hfill
	\subfigure[PAC and CPAC comparison with power constraint.]{\includegraphics[width=0.49\linewidth]{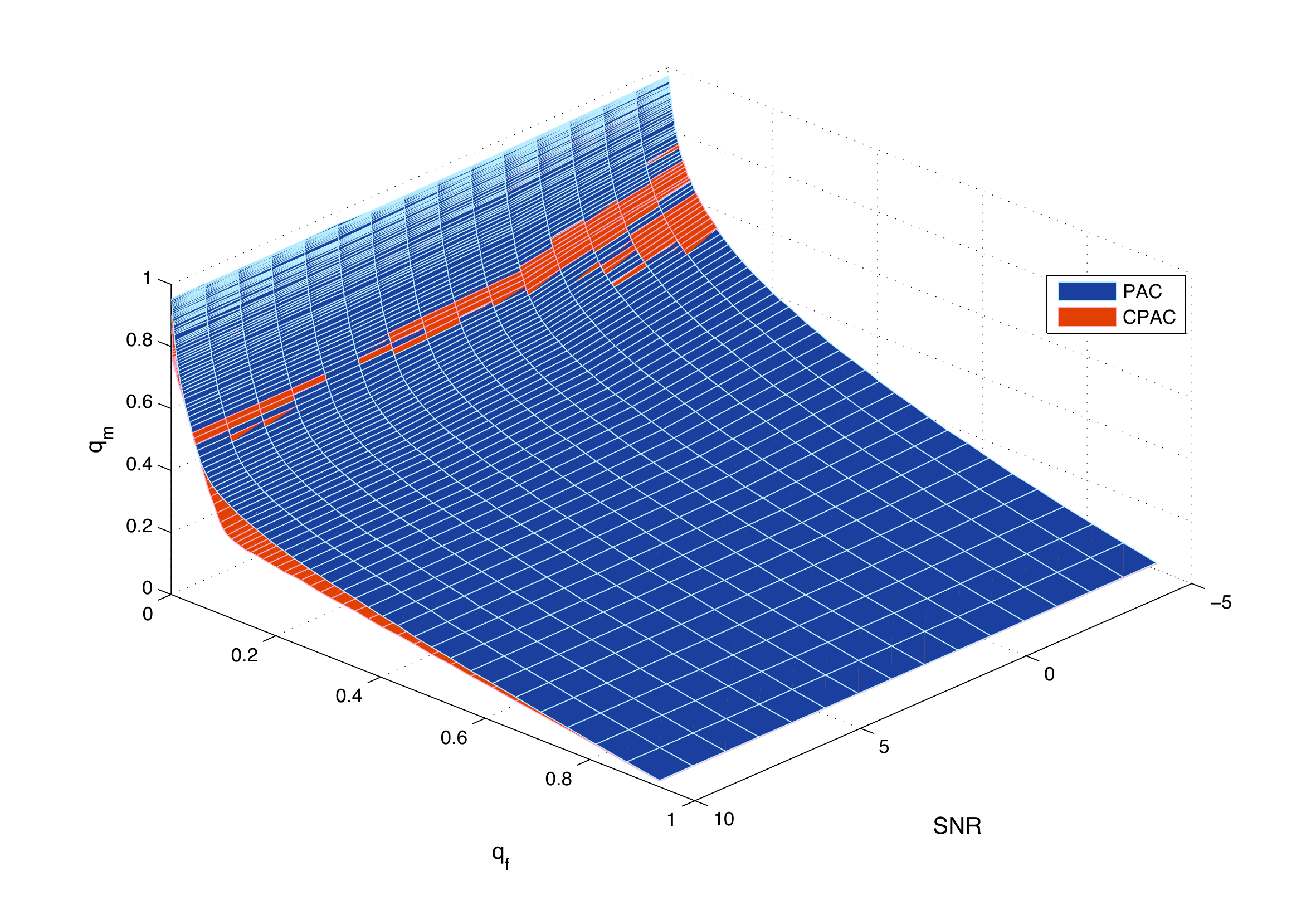}\label{fig:pPAC_pCPAC_N2}}
	\hfill
	\subfigure[PAC and CPAC comparison with energy constraint.]{\includegraphics[width=0.49\linewidth]{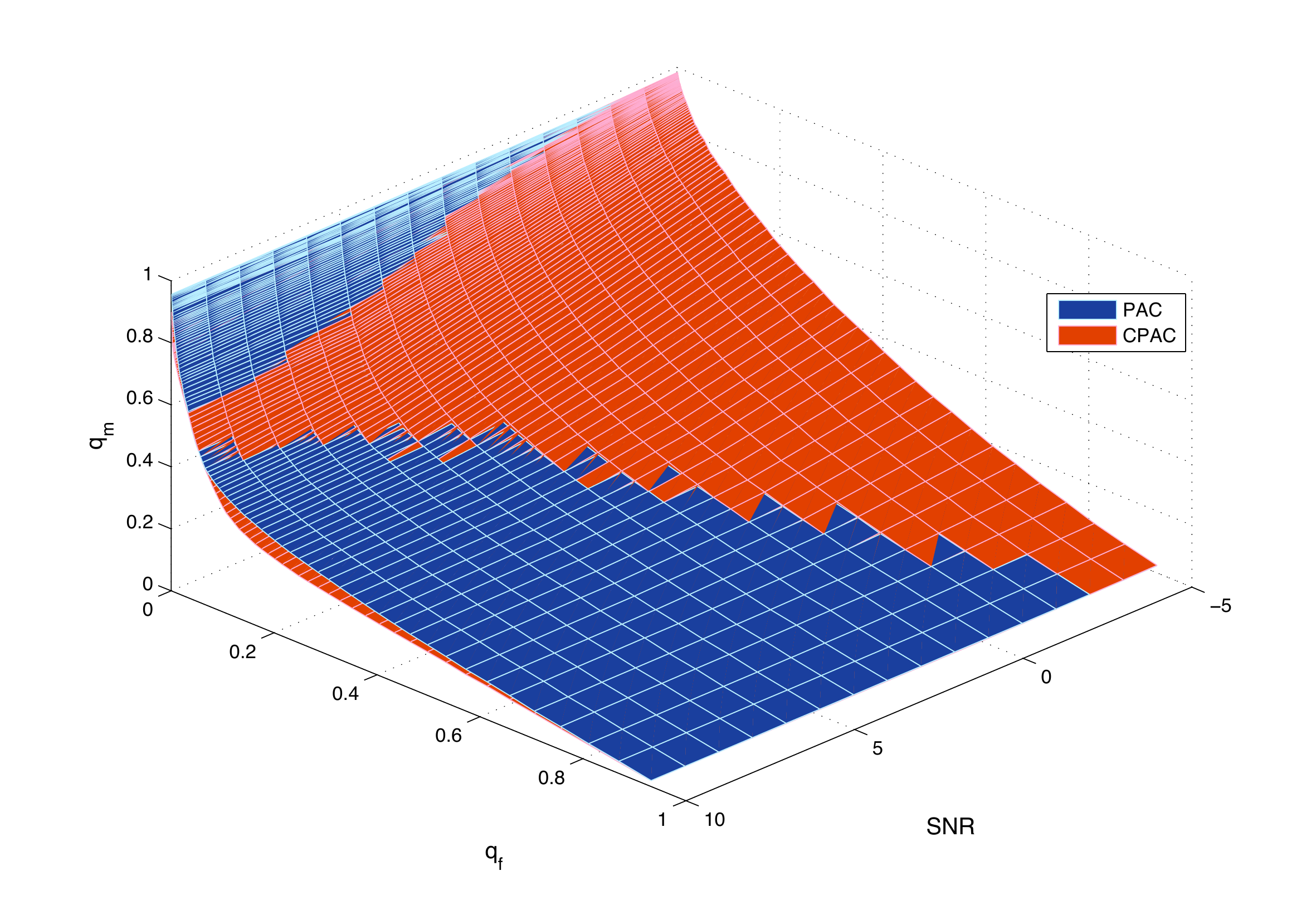}\label{fig:ePAC_eCPAC_N2}}
	\caption{3D-CROC for the four protocols in the case of $K=2$ SUs and $N=2$ receive antennas at the DFC.}\label{fig:3D_CROC_N2}
\end{figure*}

In  this section we compare the performance of the four considered protocols through the CROC obtained both via analytical results from Sec.~\ref{sec:decisionfusion} as well as from Monte Carlo simulations.
More specifically, we assume that the local performances of the sensing protocol at SU location are $p_f=0.05$ and $p_d=0.5$.
We have considered scenarios in which the number of SUs is $K\in\{2,~3\}$, the number of receive antennas at the DFC is $N\in\{1,~2,~4\}$ and the SNR $\in\{-5,~0,~5,~10\}$~dB, in both cases with unitary power and energy constraints.
However, due to space limitations, only a subset of the results are shown.
Quasi-static scenarios have been simulated with channel coefficients generated according to a Rayleigh fading model with unitary mean power.
Performance for Monte Carlo simulations were averaged over $10^5$ realizations for each considered value of the threshold.
Performance for analytical results through Gauss-Chebyshev quadrature rules used $500$ nodes for each considered value of the threshold \cite{biglieri}.

The benchmark for performance evaluation is the (lower) {\em observation bound} \cite{ciuonzo}, i.e. the performance achieved by the optimum test in the ideal case that the reporting channel is perfect,
\begin{align}
	q_f&=\sum_{\nu=g}^{K}\binom{K}{\nu}p_f^\nu(1-p_f)^{K-\nu} \;,\\
	q_m&=\sum_{\nu=0}^{g-1}\binom{K}{\nu}p_d^\nu(1-p_d)^{K-\nu} \;,
\end{align}
where $g$ is a discrete threshold.
Analytical and simulation results will show how cooperative protocols bend CROC curves toward the observation bound much more than corresponding CROC curves of non-cooperative protocols.

Fig.~\ref{fig:CROC_N2} compares the performance of the four protocols showing the CROC in the case of $N=2$ receive antennas at the DFC for SNR $\in\{0,~10\}$~dB.
Numerical simulations (shown with different markers for each protocol: $\times$ for MAC, $+$ for PAC, $\Diamond$ for CMAC and $\square$ for CPAC) confirm the correctness of the analytical results (shown with solid and dashed lines for SNR$=0$ dB and SNR$=10$ dB, respectively: red for MAC, blue for PAC, magenta for CMAC and cyan for CPAC).
Obviously, all protocols, for both power and energy constraint, benefit from an increase of the SNR, however the effect is more pronounced in the case of cooperative protocols.
Also, it is apparent how, in the case of power constraint, user cooperation improves the performance in terms of CROC in a wide range of SNR values, while in the case of energy constraint user cooperation is beneficial only for larger SNR values, while at low SNR values introduces performance degradation.
As an example, from Fig.~\ref{fig:pCROC_N2_K2}, at the operation point $(q_f,q_m)\approx(0.1,0.4)$ the curves ``MAC - SNR=0~dB'' and ``PAC - SNR=0~dB'' intersect with ``CMAC - SNR=10~dB''  and ``CPAC - SNR=10~dB'', i.e. user cooperation provides an SNR gain of $10$ dB.

Fig.~\ref{fig:CROC_10dB} compares the performance of the two protocols showing the CROC in the case of SNR $\in\{0,~10\}$~dB for the $N\in\{1,~4\}$.
Again, numerical simulations (shown with different markers for each protocol: $\times$ for MAC, $+$ for PAC, $\Diamond$ for CMAC and $\square$ for CPAC) and analytical results (shown with solid and dashed lines for $N=1$ receive antenna and $N=4$ receive antennas, respectively: red for MAC, blue for PAC, magenta for CMAC and cyan for CPAC) exhibit a good match.
As an example, from Fig.~\ref{fig:pCROC_10dB_K2}, at the operation point $(q_f,q_m)\approx(0.11,0.33)$ (resp. $(q_f,q_m)\approx(0.21,0.24)$) the curves ``MAC - $N=4$'' and ``CMAC - $N=1$'' (resp. ``PAC - $N=4$'' and ``CPAC - $N=1$'' ) intersect, i.e. user cooperation allows for reduction of the number of receive antennas at the DFC from $N=4$ to $N=1$.

Fig.~\ref{fig:3D_CROC_N2} shows three-dimensional CROCs, i.e. $q_m$ as a function of $q_f$ and SNR in the case of $N=2$ receive antennas at the DFC.
Though a numerical comparison is difficult to be done, it clearly shows how in the case of power constraint cooperative protocols outperform the corresponding noncooperative protocols over all the considered SNR range.
In the case of energy constraint, the SNR threshold above which user cooperation becomes beneficial is practically found between $0$~dB and $5$~dB.
The more critical behavior in the case of energy constraint is motivated by the fact that cooperation requires each SU spending energy for being both a source and a relay, the last phase being an additional transmission phase with respect to the non-cooperative case.
If the energy budget is very limited both the source-phase and the relay-phase transmissions undergo extremely low SNR conditions thus destroying the benefit of cooperation and making performance even worse than the case without cooperation. 

The main point of this paper is that user cooperation, a technique recently developed and analyzed thoroughly in the context of multiuser communications, is appealing also in the context of spectrum sensing through MIMO decision fusion.
Once the desired operation point has been selected, the beneficial effect of user cooperation is twofold as it allows on one side to reduce the necessary SNR, on the other side to reduce the number of received antennas at the DFC (and the corresponding computational complexity).
Also, from all the considered figures, it is apparent that the beneficial effect of orthogonal (non-interfering) transmissions is significant in the case of non-cooperative protocols, while becomes negligible in the case of cooperative protocols.
Such claims are confirmed both through simulations and analytical results.

Many issues are to be investigated such as: the overhead coming from grouping SUs into cooperative groups, the impact of errors on the cooperative channels, the extensions to other cooperative protocols than decode-and-forward, the impact of channel estimation errors at the DFC.
It is worth noticing that spectrum sensing is beneficial when the average duration of the event ``PU is silent'' is much longer then the time needed for sensing. 
User cooperation makes the sensing process longer in time, then a negative effect is that it reduces the scenarios in which it is applicable. 
An analysis of the achievable throughput depending on the statistics of the PU activity is out of the scope of this paper.
A final remark must be given with reference to the choice of the architecture.
It is worth noticing that cooperative protocols may be easily implemented in a distributed manner without the need of the fusion center and with each SUs implementing locally its fusion rule.
This interesting scenario opens different challenges: the computational capabilities at SU location;  the possibility to employ multiple antennas at the SU location, the presence of different final decisions in the system, etc. 
However, centralized architectures could be still preferred because the fusion center is an ideal candidate to resolve other issues such as synchronization and coordination SUs.

%% file: Cooperative_Spectrum_Sensing---secAPP.tex
\appendix

From the total probability theorem:
\begin{align}
	\Phi_{-\Lambda}(s|{\cal H}_i)&=\sum_{\x\in\{-1,+1\}^K} {\Phi_{-\Lambda}(s|\x)\Pr(\x|{\cal H}_i)} \label{eq:tptmgf}\;.
\end{align}
It is worth noticing that the right term in Eq.~(\ref{eq:tptmgf}) would require $2^K$ summations, however, it will be clear in the following that the needed summations are $K+1$ as the dependence on $\x$ is through the number of sensors that transmit $+1$ defined as follows
\begin{align}
	\nu=\frac{1}{2}\left(K+\sum_{k=1}^Kx_k\right) \;.
\end{align}

Denoting the following vectors and matrices:
\begin{align}
	\h_n[\ell]&=\H_{eq}^t\i_{M}^{((\ell-1)N+n)} \;,\\
	\v_n[\ell]&=\left(y_n[\ell], \h_n[\ell]^t\1_K\right)^t \;,\\
	\v_n&=\left(\v_n[1]^t,\ldots,\v_n[L]^t\right)^t \;,\\
	\F&=\left(\begin{array}{cc} 0 & -1/2 \\ -1/2 & 0\end{array}\right) \;,\\
	\R(\x)&=\Ex\left\{(\v_n|\x)(\v_n|\x)^\dag\right\} \;,\\ 
	\Q(\x)&=\R(\x)\cdot(\I_L\otimes\F) \;,
\end{align}
it is straightforward to show that
\begin{align}
	-\Lambda=\sum_{n=1}^N\sum_{\ell=1}^L \v_n[\ell]^\dag\F\v_n[\ell] 
	=\sum_{n=1}^N \v_n^\dag (\I_L\otimes\F)\v_n \;,
\end{align}
being the sum of Hermitian quadratic forms of circularly complex Gaussian vectors, and then \cite{schwarz}
\begin{align}
	\Phi_{-\Lambda}(s|\x)&=\left(\frac{1}{\det\left(\I_{2L}+s\Q(\x)\right)}\right)^{N} \label{eq:condmgf}\;,
\end{align}
where we exploited the fact that $\{\v_n|\x\}_{n=1}^N$ are i.i.d. vectors.

In the following, we compute $\det\left(\I_{2L}+s\Q(\x)\right)$ for each protocol in order to evaluate Eq.~(\ref{eq:condmgf}).
We denote $(\cdot)_N$ the modulo operation over the set of integers $\{1,\ldots, N\}$ (i.e. the modulo$-N$ operator except from replacing $0$ with $N$).
The cyclic-permutation matrix of size $N$ is denoted $\P_N$, i.e.
\begin{align}
	\P_N&=\left(\begin{array}{cc} 
	\0_{N-1}^t & 1 \\ 
	\I_{N-1} & \0_{N-1} \\ 
	\end{array}\right) \;.
\end{align}	 
Also, we define the following matrices and scalars:
\begin{align}
	\A&=\left(\begin{array}{cc} 
	K+\sigma_w^2 & \sum_{k=1}^K x_k \\ 
	\sum_{k=1}^{K} x_k & K 
	\end{array}\right) \;,\\
	\tilde{\A}&=\I_2+s\A\F \;,\\
	a&=\det(\tilde{\A})=1+(K-2\nu)s+\left(\nu(\nu-K)-\frac{K}{4}\sigma_w^2\right)s^2
\end{align}
\begin{align}
	\B_k&=\left(\begin{array}{cc} 
	1+\sigma_w^2 & x_k \\ 
	x_k & 1 
	\end{array}\right) \;,\\
	\tilde{\B}_k&=\I_2+s\B_k\F \;,\\
	b_k&=\det(\tilde{\B}_k)=1-x_ks-\left(\frac{\sigma_w^2}{4}\right)s^2 \;,
\end{align}
\begin{align}	 
	\C_{k,i}&=\left(\begin{array}{cc} 
	x_kx_i & x_k \\ 
	x_i & 1 
	\end{array}\right) \;,\\
	\D_i&=\left(\begin{array}{cc} 
	\sum_{k=1}^K x_kx_{(k-i)_K} & 2\nu-K \\ 
	2\nu-K & K 
	\end{array}\right) \;,
\end{align}
\begin{align}	
	\E_{k,i}&=s\C_{k,i}\F\tilde{\B}_i^{-1} \;,\\
	\G_{k,m,i}&=s\E_{k,m}\C_{m,i}\F \;,\\
	\M_{k,i}&=\sum_{m=1}^K\G_{(m-k)_{K},m,(m-i)_{M}} \;,
\end{align}
\begin{align}
	\U&={\rm diag}\left(\B_1,\ldots,\B_K\right) \;,\\
	\tilde{\U}&={\rm diag}\left(\tilde{\B}_1,\ldots,\tilde{\B}_K\right) \;,\\
	u&=\det(\tilde{\U})=\prod_{k=1}^K b_k \;,
\end{align}
\begin{align}	 
	\V&=\left(\begin{array}{cccc} 
	\C_{1,K} & \C_{1,K-1} & \cdots & \C_{1,2} \\ 
	\C_{2,1} & \C_{2,K} & \cdots & \C_{2,3} \\ 
	\vdots & \vdots & \ddots & \vdots \\ 
	\C_{K,K-1} & \C_{K,K-2} & \cdots & \C_{K,1} 
	\end{array}\right) \;,
\end{align}
\begin{align}	 
	\W&=\left(\begin{array}{cccc} 
	\A & \D_{1} & \cdots & \D_{K-2} \\ 
	\D_{1} & \A & \cdots & \D_{K-3} \\ 
	\vdots & \vdots & \ddots & \vdots \\ 
	\D_{K-2} & \D_{K-3} & \cdots & \A 
	\end{array}\right) \;,
\end{align}
\begin{align}	 
	\tilde{\W}&=\left(\begin{array}{cccc} 
	\tilde{\A} & s\D_{1}\F & \cdots & s\D_{K-2}\F \\ 
	s\D_{1}\F & \tilde{\A} & \cdots & s\D_{K-3}\F \\ 
	\vdots & \vdots & \ddots & \vdots \\ 
	s\D_{K-2}\F & s\D_{K-3}\F & \cdots & \tilde{\A} 
	\end{array}\right) \;,
\end{align}
\begin{align}	 
	\tilde{\M}&=\left(\begin{array}{ccc} 
	\M_{1,1} & \cdots & \M_{1,K-1} \\ 
	\vdots & \ddots & \vdots \\ 
	\M_{K-1,1} & \cdots & \M_{K-1,K-1} 
	\end{array}\right) \;,
\end{align}
\begin{align}
	\S_{k,i}&={\rm diag}\left(\C_{1,(1+k-i)_K},\C_{2,(2+k-i)_K},\right. \nonumber\\
	&~\left. \ldots,\C_{K,(K+k-i)_K}\right) \;,\qquad k<i\;,\\
	\Z_{k,i}&=\begin{cases}
	\S_{k,i}\cdot\left(\P_K^{i-k}\otimes\I_2\right) \qquad {k<i}\cr
	\left(\P_K^{k-i}\otimes\I_2\right)^t \cdot \S_{i,k}^t \qquad {k>i}
	\end{cases}\;.
\end{align}

\subsection{Derivation of the MGF for the MAC protocol}
From the equivalent channel in Eq.~(\ref{eq:mac_set}) we obtain
\begin{align}
	\R(\x)&=\A \;,\\
	\Q(\x)&=\A\F \;.
\end{align}
from which we get
\begin{align}
	\det\left(\I_2+s\Q(\x)\right)&=a \label{eq:det_mac}\;.
\end{align}
Replacing Eq.~(\ref{eq:det_mac}) into (\ref{eq:condmgf}) and then using (\ref{eq:tptmgf}), we get (\ref{eq:mgf_mac_0_K}).

\subsection{Derivation of the MGF for the PAC protocol}
From the equivalent channel in Eq.~(\ref{eq:pac_set}) we obtain
\begin{align}
	\R(\x)&={\rm diag}\left(\B_1,\ldots,\B_K\right) \;,\\
	\Q(\x)&={\rm diag}\left(\B_1\F,\ldots,\B_K\F\right) \;.
\end{align}
from which we get
\begin{align}
	\det\left(\I_{2K}+s\Q(\x)\right)&=u \label{eq:det_pac}\;.
\end{align}
Replacing Eq.~(\ref{eq:det_pac}) into (\ref{eq:condmgf}) and then using (\ref{eq:tptmgf}), we get (\ref{eq:mgf_pac_0_K}).

\subsection{Derivation of the MGF for the CMAC protocol}
From the equivalent channel in Eq.~(\ref{eq:cmac_set}) we obtain
\begin{align}
	\R(\x)&=\left(\begin{array}{cc} 
	\U & \V \\ 
	\V^t & \W 
	\end{array}\right) \;,\\
	\Q(\x)&=\left(\begin{array}{cc} 
	\U\cdot(\I_{K}\otimes\F) & \V\cdot(\I_{K-1}\otimes\F) \\ 
	\V^t\cdot(\I_{K}\otimes\F) & \W\cdot(\I_{K-1}\otimes\F) 
	\end{array}\right) \;.
\end{align}
from which we get
\begin{align}
	\det\left(\I_{4K-2}+s\Q(\x)\right)&=u\det\left(\tilde{\W}-\tilde{\M}\right)  \nonumber\\
	&= u\det(\tilde{\W})\det\left(\I_{2K-2}-\tilde{\M}\tilde{\W}^{-1}\right)  \label{eq:det_cmac}\;.
\end{align}
where we have applied Sylvester's determinant theorem.

More specifically, in the case with $K=2$ SUs Eq.~(\ref{eq:det_cmac}) provides
\begin{align}
	\det\left(\I_{6}+s\Q(\x)\right)=& 1+4(1-\nu)s+\left(3(\nu-1)^2-\sigma_w^2\right)s^2 \nonumber\\
	&+\frac{3}{2}(\nu-1)\sigma_w^2s^3+\frac{3}{16}\sigma_w^4s^4 \label{eq:det_cmac_k2}\;.
\end{align}
Replacing Eq.~(\ref{eq:det_cmac_k2}) into (\ref{eq:condmgf}) and then using (\ref{eq:tptmgf}), we get (\ref{eq:mgf_cmac_0_k2}).
Differently, in the case with $K=3$ SUs Eq.~(\ref{eq:det_cmac}) provides
\begin{align}
	&\det\left(\I_{10}+s\Q(\x)\right)= 1+3(3-2\nu)s+\left(7\nu^2-21\nu+15\right.  \nonumber\\
	&\left.-\frac{9}{4}\sigma_w^2\right)s^2+\left(\frac{5\sigma_w^2-3}{2}(2\nu-3)+\frac{5}{2}(-1)^\nu\right)s^3\nonumber\\
	&+\frac{\sigma_w^2}{4}\left(\frac{15}{4}\sigma_w^2-21-11\nu^2+33\nu\right)s^4-\frac{7}{16}\sigma_w^4(2\nu-3)s^5\nonumber\\
	&-\frac{7}{64}\sigma_w^6s^6
	\label{eq:det_cmac_k3}\;.
\end{align}
Replacing Eq.~(\ref{eq:det_cmac_k3}) into (\ref{eq:condmgf}) and then using (\ref{eq:tptmgf}), we get (\ref{eq:mgf_cmac_0_k3}).

\subsection{Derivation of the MGF for the CPAC protocol}
From the equivalent channel in Eq.~(\ref{eq:cpac_set}) we obtain
\begin{align}
	\R(\x)&=\left(\begin{array}{cccc} 
	\U & \Z_{1,2} & \cdots & \Z_{1,K} \\ 
	\Z_{2,1} & \U & \cdots & \Z_{2,K} \\ 
	\vdots & \vdots & \ddots & \vdots \\ 
	\Z_{K,1} & \Z_{K,2} & \cdots & \U
	\end{array}\right) \;,\\
	\Q(\x)&=\left(\begin{array}{ccc} 
	\U\cdot(\I_{K}\otimes\F) & \cdots & \Z_{1,K}\cdot(\I_{K}\otimes\F) \\ 
	\Z_{2,1}\cdot(\I_{K}\otimes\F) & \cdots & \Z_{2,K}\cdot(\I_{K}\otimes\F) \\ 
	\vdots & \ddots & \vdots \\ 
	\Z_{K,1}\cdot(\I_{K}\otimes\F) & \cdots & \U\cdot(\I_{K}\otimes\F)
	\end{array}\right) \;.
\end{align}
from which we get
\begin{align}
	\det\left(\I_{2K^2}+s\Q(\x)\right)&=
	\left(1-\left(\sum_{k=1}^Kx_k\right)s-\frac{K}{4}\sigma_w^2s^2\right)^K \label{eq:det_cpac}\;.
\end{align}
Replacing Eq.~(\ref{eq:det_cpac}) into (\ref{eq:condmgf}) and then using (\ref{eq:tptmgf}) we get (\ref{eq:mgf_cpac_0_K}).

%% file: Cooperative_Spectrum_Sensing---main.bbl
\begin{thebibliography}{1}

\bibitem{mitola}
J.~Mitola~III and G.~Q.~Maguire,~Jr., 
``Cognitive radio: making software radios more personal,''
\emph{IEEE Personal Commun.}, 
vol.~6, no.~4, pp.~13--18, Aug.~1999.

\bibitem{haykin}
S.~Haykin, 
``Cognitive radio: brain-empowered wireless communications,''
\emph{IEEE J. Sel. Areas Commun.}, 
vol.~23, no.~2, pp.~201--220, Feb.~2005.

\bibitem{akyildiz2006}
I.~F.~Akyildiz, W.~Y.~Lee, M.~C.~Vuran, S.~Mohanty, 
``NeXt generation / dynamic spectrum access / cognitive radio wireless networks: a survey,'' 
\emph{Comp. Net. (Elsevier)}, 
vol.~50, no.~13, pp.~2127--2159, Sep.~2006.

\bibitem{akyildiz2011}
I.~F.~Akyildiz, B.~F.~Lo, and R.~Balakrishnan, 
``Cooperative spectrum sensing in cognitive radio networks: a survey,'' 
\emph{Physical Commun. (Elsevier)}, 
vol.~4, no.~1, pp.~40--62, Mar.~2011.

\bibitem{axell}
E.~Axell, G.~Leus, E.~G.~Larsson, and H.~V.~Poor,
``Spectrum sensing for cognitive radio: state-of-the-art and recent advances,'' 
\emph{IEEE Signal Process. Mag.}, 
vol.~29, no.~3, pp.~101--116 , May.~2012.

\bibitem{urkowitz}
H.~Urkowitz, 
``Energy detection of unknown deterministic signals,''
\emph{Proc.~IEEE}, 
vol.~55, no.~4, pp.~523--531, Apr.~1967.

\bibitem{digham}
F.~F.~Digham, M.~S.~Alouini, and M.~K.~Simon, 
``On the energy detection of unknown signals over fading channels,''
\emph{IEEE Trans. Commun.}, 
vol.~55, no.~1, pp.~21--24, Jan.~2007.

\bibitem{lunden}
J.~Lunden, V.~Koivunen, A.~Huttunen, and H.~V.~Poor, 
``Collaborative cyclostationary spectrum sensing for cognitive radio systems,'' 
\emph{IEEE Trans. Signal Process.}, 
vol.~57, pp.~4182--4195, Nov.~2009.

\bibitem{mishra}
S.~M.~Mishra, A.~Sahai, and R.~Brodersen, 
``Cooperative sensing among cognitive radios,''
in \emph{Proc. IEEE Int.. Conf. Commun. (ICC)}, 
vol.~4, pp.~1658--1663 , Jun.~2006.

\bibitem{unnikrishnan}
J.~Unnikrishnan, V.~V.~Veeravalli, 
``Collaborative sensing for primary detection in cognitive radio,'' 
\emph{IEEE Sel. Topics Signal Process.}, 
vol.~2, no.~1, pp.~18--27, Feb.~2008.

\bibitem{quan}
Z.~Quan, S.~Cui, and A.~H.~Sayed, 
``Optimal linear cooperation for spectrum sensing in cognitive radio networks,'' 
\emph{IEEE J. Sel. Topics Signal Process.}, 
vol.~2, no.~1, pp.~28--40, Feb.~2008.

\bibitem{letaief}
W.~Zhang and K.~B.~Letaief, 
``Cooperative spectrum sensing with transmit and relay diversity in cognitive radio networks,'' 
\emph{IEEE Trans. Wireless Commun.}, 
vol.~7, no.~12, pp.~4761--4766, Dec.~2008.

\bibitem{li}
Z.~Li, F.~R.~Yu, and M.~Huang, 
``A distributed consensus-based cooperative spectrum-sensing scheme in cognitive radios,''
\emph{IEEE Trans. Vehic. Tech.}, 
vol.~59, no.~1, pp.~383--393, Jan.~2010.

\bibitem{ma}
J.~Ma, G.~Zhao, and Y.~(G.)~Li, 
``Soft combination and detection for cooperative spectrum sensing in cognitive radio networks,'' 
\emph{IEEE Trans. Wireless Commun.}, 
vol.~7, no.~11, pp.~4502--4507, Nov.~2008.

\bibitem{zhang}
W.~Zhang, R.~K.~Mallik, and K.~B.~Letaief, 
``Optimization of cooperative spectrum sensing with energy detection in cognitive radio networks,'' 
\emph{IEEE Trans. Wireless Commun.}, 
vol.~8, no.~12, pp.~5761--5766, Dec.~2009.

\bibitem{saad}
W.~Saad, Z.~Han, T.~Basar, M.~Debbah, and A.~Hj{\o}rungnes,
``Coalition formation games for collaborative spectrum sensing,''
\emph{IEEE Trans. Vehic. Tech.}, 
vol.~60, no.~1, pp.~276--297, Jan.~2011.

\bibitem{goldsmith}
A.~Goldsmith,
\emph{Wireless Communications.}
Cambridge University Press, 2005.

\bibitem{sendonaris}
A.~Sendonaris, E.~Erkip, and B.~Aazhang, 
``User cooperation diversity -- Part I: System description,'' and ``Part II: Implementation aspects and performance analysis,'' 
\emph{IEEE Trans. Commun.}, 
vol.~51, no.~11, pp.~1927--1948, Nov.~2003.

\bibitem{laneman2003}
J.~N.~Laneman and G.~W.~Wornell, 
``Distributed space-time coding protocols for exploiting cooperative diversity in wireless networks,'' 
\emph{IEEE Trans. Inform. Theory}, 
vol.~49, no.~10, pp.~2415--2425, Oct.~2003.

\bibitem{laneman2004}
J.~N.~Laneman, D.~N.~C.~Tse, and G.~W.~Wornell, 
``Cooperative diversity in wireless networks: Efficient protocols and outage behavior,'' 
\emph{IEEE Trans. Inform. Theory}, 
vol.~50, no.~12, pp.~3062--3080, Dec.~2004.

\bibitem{nosratinia2004}
M.~Janani, A.~Hedayat, T.~E.~Hunter, and A.~Nosratinia, 
``Coded cooperation in wireless communications: space-time transmission and iterative decoding,''
\emph{IEEE Trans. Signal Process.}, 
vol.~52, no.~2, pp.~362--370, Feb.~2004.

\bibitem{nosratinia2006}
T.~E.~Hunter and A.~Nosratinia, 
``Diversity through coded cooperation,''
\emph{IEEE Trans. Wireless Commun.},
 vol.~5, no.~2, pp.~283--289, Feb.~2006.

\bibitem{larsson}
E.~G.~Larsson and B.~R.~Vojcic, 
``Cooperative transmit diversity based on superposition modulation,''
\emph{IEEE Commun. Lett.} 
vol.~9, no.~9, pp. 778--780, Sep.~2005.

\bibitem{costello}
L.~Xiao, T.~E.~Fuja, J.~Kliewer, and D.~G.~Costello, Jr., 
``A network coding approach to cooperative diversity,'' 
\emph{IEEE Trans. Inform. Theory}, 
vol.~53, no.~10, pp.~3714--3722, Oct.~2007.

\bibitem{salvorossi2007}
P.~Salvo~Rossi, A.~P.~Petropulu, F.~Palmieri, and G.~Iannello, 
``Distributed linear block coding for cooperative wireless communications,'' 
\emph{IEEE Signal Process. Lett.}, 
vol.~14, no.~10, pp.~673--676, Oct.~2007.

\bibitem{salvorossi2010}
P.~Salvo~Rossi, 
``On the performance of cooperative systems using distributed linear block coding,'' 
\emph{Physical Commun. (Elsevier)}, 
vol.~3, no.~2, pp.~81--86, Jun.~2010.

\bibitem{ganesan}
G.~Ganesan and Y.~(G.)~Li, 
``Cooperative spectrum sensing in cognitive radio -- Part I: two user networks,'' and ``Part II: multiuser networks,'' 
\emph{IEEE Trans. Wireless Commun.}, 
vol.~6, no.~6, pp.~2204--2222, Jun.~2007.

\bibitem{chen}
B.~Chen, L.~Tong, and P.~K.~Varshney, 
``Channel-aware distributed detection in wireless sensor networks,'' 
\emph{IEEE Signal Process. Mag.},
vol.~23, no.~4, pp.~16--26, Jul.~2006.

\bibitem{banavar}
M.~K.~Banavar, A.~D.~Smith, C.~Tepedelenlioglu, and A.~Spanias, 
``On the effectiveness of multiple antennas in distributed detection over fading MACs,'' 
\emph{IEEE Trans. Wireless Commun.}, 
vol.~11, no.~5, pp.~1744--1752, May~2012.

\bibitem{ciuonzo}
D.~Ciuonzo, G.~Romano, and P.~Salvo~Rossi,  
``Channel-aware decision fusion in distributed MIMO wireless sensor networks: decode-and-fuse vs. decode-then-fuse,''
\emph{IEEE Trans. Wireless Commun.}, 
vol.~11, no.~8, pp.~2976--2985, Aug.~2012.

\bibitem{ciuonzon}
D.~Ciuonzo, G.~Romano, and P.~Salvo~Rossi,  
``Optimality of received energy in decision fusion over a Rayleigh fading diversity MAC with non-identical sensors,''
\emph{IEEE Trans. Signal Process.}, 
vol.~61, no.~1, pp.~22--27, Jan.~2013.

\bibitem{suzuki}
K.~Umebayashi, J.~J.~Lehtom\"aki, T.~Yazawa, and Y.~Suzuki,
``Efficient decision fusion for cooperative spectrum sensing based on OR-rule,''
\emph{IEEE Trans. Wireless Commun.}, 
vol.~11, no.~7, pp.~2585--2595, Jul.~2012.

\bibitem{peh2011}
E.~C.~Y.~Peh, Y.-C.~Liang, Y.~L.~Guan, and Y.~Zeng,  
``Power control in cognitive radios under cooperative and non-cooperative spectrum sensing,''
\emph{IEEE Trans. Wireless Commun.}, 
vol.~10, no.~12, pp.~4238--4248, Dec.~2011.

\bibitem{salvorossi2011}
P.~Salvo~Rossi and G.~M.~Kraidy,  
``Iterative multiuser detection for cooperative MIMO systems over quasi-static channels,''
\emph{IEEE Trans. Wireless Commun.}, 
vol.~10, no.~11, pp.~3638--3643, Nov.~2011.

\bibitem{schober}
A.~Lei and R.~Schober, 
``Coherent max-log decision fusion in wireless sensor networks,'' 
\emph{IEEE Trans. Commun.}, 
vol.~58, no.~5, pp.~1327--1332, May~2010. 

\bibitem{peh2008}
Y.-C.~Liang, Y.~Zeng, E.~C.~Y.~Peh, and A.~T.~Hoang,  
``Sensing-throughput tradeoff for cognitive radio networks,''
\emph{IEEE Trans. Wireless Commun.}, 
vol.~7, no.~4, pp.~1326--1337, Apr.~2008.

\bibitem{peh2009}
E.~C.~Y.~Peh, Y.-C.~Liang, Y.~L.~Guan, and Y.~Zeng, 
``Optimization of cooperative sensing in cognitive radio networks: a sensing-throughput tradeoff view,''
\emph{IEEE Trans. Veh. Technol.}, 
vol.~58, no.~9, pp.~5294--5299, Nov.~2009.

\bibitem{ciuonzo2}
D.~Ciuonzo, G.~Romano, and P.~Salvo~Rossi,  
``Performance analysis and design of maximum ratio combining in channel-aware MIMO decision fusion,''
\emph{IEEE Trans. Wireless Commun.}, 
in press.

\bibitem{kay}
S.~M.~Kay, 
\emph{Fundamentals of Statistical Signal Processing, Volume 2: Detection Theory.}
Prentice Hall PTR, 1998.

\bibitem{biglieri}
E.~Biglieri, G.~Caire, G.~Taricco, and J.~Ventura-Traveset, 
``Computing error probabilities over fading channels: a unified approach,'' 
\emph{Eur. Trans. Telecommun.}, vol.~9, no.~1, pp.~15--25, Jan.~1998.

\bibitem{annamalai}
A.~Annamalai, C.~Tellambura, and V.~K. ~Bhargava, 
``Efficient computation of MRC diversity performance in Nakagami fading channel with arbitrary parameters,'' 
\emph{Electron. Lett.}, vol.~34, no.~12, pp.~1189--1190, Jun.~1998.

\bibitem{schwarz} 
M.~Schwarz, W.~R.~Bennet, and S.~Stein, 
\emph{Communication Systems and Techniques.} 
New York: McGraw-Hill, 1966.

\end{thebibliography}
